\documentclass[12pt, draftclsnofoot, onecolumn]{IEEEtran}
\usepackage{amsthm}
\usepackage{multirow}
\usepackage{color}

\usepackage[T1]{fontenc}
\usepackage{algorithm}
\usepackage{algorithmic}

\ifCLASSINFOpdf
\else
\fi
\usepackage{amsmath}
\usepackage{amsmath,bm}
\usepackage{graphicx}
\usepackage{amsmath,amssymb,amsthm,mathrsfs,amsfonts,dsfont}
\usepackage{subfigure}
\usepackage{array}
\begin{document}
\title{High-Resolution Angle Tracking for\\ Mobile Wideband Millimeter-Wave Systems with Antenna Array Calibration}

\author{Dalin Zhu,
        Junil Choi,
        Qian Cheng,
        Weimin Xiao,
        and~Robert~W.~Heath~Jr.
\thanks{Dalin Zhu and Robert W. Heath Jr. are with the Department
of Electrical and Computer Engineering, The University of Texas at Austin, Austin,
TX, 78712 USA, e-mail: \{dalin.zhu, rheath\}@utexas.edu.

Junil Choi is with the Department of Electrical Engineering, Pohang University of Science and Technology (POSTECH), Pohang, Gyeongbuk 37673 Korea, e-mail: junil@postech.ac.kr.

Qian Cheng and Weimin Xiao are with the Wireless Research and Standards Department, Huawei R\&D USA, Rolling Meadows, IL, 60008 USA, e-mail: \{q.cheng, weimin.xiao\}@huawei.com

This work was supported in part by a gift from Huawei Technologies, in part by the National Science Foundation under Grant No. 1702800, and in part by the Institute for Information and Communications Technology Promotion (IITP) grant funded by the Korean Government (MSIT) (No.2018(2016-0-00123), Development of Integer-Forcing MIMO Transceivers for 5G and Beyond Mobile Communication Systems).}}

\maketitle

\begin{abstract}
Millimeter-wave (mmWave) systems use directional beams to support high-rate data communications. Small misalignment between the transmit and receive beams (e.g., due to the mobility) can result in significant drop of the received signal quality especially in line-of-sight communication channels. In this paper, we propose and evaluate high-resolution angle tracking strategies for wideband mmWave systems with mobility. We custom design pairs of auxiliary beams as the tracking beams, and use them to capture the angle variations, towards which the steering directions of the data beams are adjusted. Different from conventional beam tracking designs, the proposed framework neither depends on the angle variation model nor requires an on-grid assumption. For practical implementation of the proposed methods, we examine the impact of the array calibration errors on the auxiliary beam pair design. Numerical results reveal that by employing the proposed methods, good angle tracking performance can be achieved under various antenna array configurations, channel models, and mobility conditions.
\end{abstract}


%
\IEEEpeerreviewmaketitle

\allowdisplaybreaks

\section{Introduction}

The small array form factor at millimeter-wave (mmWave) frequencies enables the use of large antenna arrays to generate highly directional beams. This allows array gain for improved received signal power and also reduces mean interference levels \cite{jerrypikhan}-\nocite{rhsp}\nocite{pchj}\nocite{fbrh}\cite{mmwavebook}. For the most benefits from beamforming, accurate channel direction information such as the channel's angle-of-departures (AoDs) and angle-of-arrivals (AoAs) is required at both the base station (BS) and user equipment (UE) sides. Further, due to the UE's mobility, slight misalignments of the transmit and receive beams with the channel's AoDs and AoAs may result in significant performance loss at mmWave frequencies in line-of-sight (LOS) communication channels. Hence, accurate beam or angle tracking designs are required to better capture the channel variations and enable reliable mmWave communications in fast-varying environments.

Grid-of-beams based beam training is the defacto approach for configuring transmit and receive beams; variations are used in IEEE 802.11ad systems \cite{ieeewlan,wang} and will be used in 5G \cite{5gcon}. Beam tracking approaches that support grid-of-beams have been developed in \cite{ieeewlan,wang,bt0,bt1}, but the performance depends on the grid resolution, leading to high complexity, tracking overhead, and access delay. In \cite{bt2,bt4}, a priori-aided angle tracking strategies were proposed. By combining the temporal variation law of the AoD and AoA of the LOS path with the sparse structure of the mmWave channels, the channels obtained during the previous time-slots are used to predict the support (the index set of non-zero elements in a sparse vector) of the channel. The time-varying parameters corresponding to the support of the channel are then tracked for the subsequent time-slots. To track the non-LOS (NLOS) paths, the classical Kalman filter can be employed by first eliminating the influence of the LOS path \cite{bt3}. In \cite{bt5}, the idea of Kalman filter was exploited as well when designing the angle tracking and abrupt change detection algorithms. In \cite{vatrack}, the extended Kalman filter was used to track the channel's AoDs and AoAs by only using the measurement of a single beam pair. The angle tracking algorithms developed in \cite{bt2}-\nocite{bt4}\nocite{bt3}\nocite{bt5}\cite{vatrack}, however, depend on specific modeling of the geometric relationship between the BS and UE and the angle variations.

In this paper, we develop high-resolution angle tracking algorithms through the auxiliary beam pair design for mobile wideband mmWave systems under the analog architecture. In the employed analog architecture, the BS uses a small number of radio frequency (RF) chains to drive a large number of antenna elements, and forms the tracking beams in the analog domain. We propose and analyze new angle tracking procedures, where the basic principles follow those in \cite{dztrans,dztrans2d} with moderate modifications based on the employed array configurations and pilot signal structures. In our previous work \cite{dztrans,dztrans2d}, we exploited the idea of auxiliary beam pair design to estimate both the narrowband and wideband mmWave channels with and without dual-polarization. The proposed approaches, however, were only applied to the angle estimation, and not specifically designed for the angle tracking. Further, in this paper, we custom design two array calibration strategies for the employed analog architecture, and characterize the impact of the array calibration errors on the proposed methods. We summarize the main contributions of the paper as follows:
\begin{itemize}
  \item We provide detailed design procedures of the proposed auxiliary beam pair-assisted angle tracking approaches in wideband mmWave systems. We propose several angle tracking design options and differentiate them in terms of tracking triggering device, feedback information, and information required at the UE side.
  \item We develop and evaluate direct and differential feedback strategies for the proposed angle tracking designs in frequency-division duplexing systems. By judiciously exploiting the structure of auxiliary beam pair, the differential feedback strategy can significantly reduce the feedback overhead.
  \item We custom design two receive combining based array calibration methods for the employed analog architecture, in which all the antenna elements are driven by a small number of RF chains. The proposed two methods are different in terms of the probing strategies of the calibration reference signals.
  \item We characterize the impact of the radiation pattern impairments on our proposed methods. We first exhibit that relatively large phase and amplitude errors would contaminate the angle tracking performances of the proposed algorithms, resulting in increased tracking error probability and reduced spectral efficiency. By using the proposed array calibration methods to compensate for the radiation pattern impairments, we show that the proposed angle tracking strategies work well even with residual calibration errors.
\end{itemize}

We organize the rest of this paper as follows. In Section II, we first describe the employed system and wideband channel models; we then illustrate the frame structure and conventional grid-of-beams based beam tracking design for mmWave systems. In Section III, we explain detailed design principles and procedures of the proposed high-resolution angle tracking strategies. In Section IV, we discuss the developed array calibration methods and their impact on the proposed angle tracking designs. In Section V, we present numerical results to validate the effectiveness of the proposed techniques. Finally, we draw our conclusions in Section VI.

\textbf{Notations}: $\bm{A}$ is a matrix; $\bm{a}$ is a vector; $a$ is a scalar; $|a|$ is the magnitude of the complex number
$a$; $(\cdot)^{\mathrm{T}}$ and $(\cdot)^{*}$ denote transpose and conjugate transpose; $\bm{I}_{N}$ is the $N\times N$ identity matrix; $\bm{1}_{M\times N}$ represents the $M\times N$ matrix whose entries are all ones; $\mathcal{N}_{\mathrm{c}}(\bm{a},\bm{A})$ is a complex Gaussian vector with mean $\bm{a}$ and covariance $\bm{A}$; $\mathbb{E}[\cdot]$ is used to denote expectation; $\otimes$ is the Kronecker product; $\mathrm{sign}(\cdot)$ extracts the sign of a real number; $\mathrm{diag}(\cdot)$ is the diagonalization operation; and $\mathrm{vec}(\cdot)$ is the matrix vectorization operation.

\section{System Model and Conventional Beam Tracking Design}
\begin{figure}
\centering
\subfigure[]{%
\includegraphics[width=4.15in]{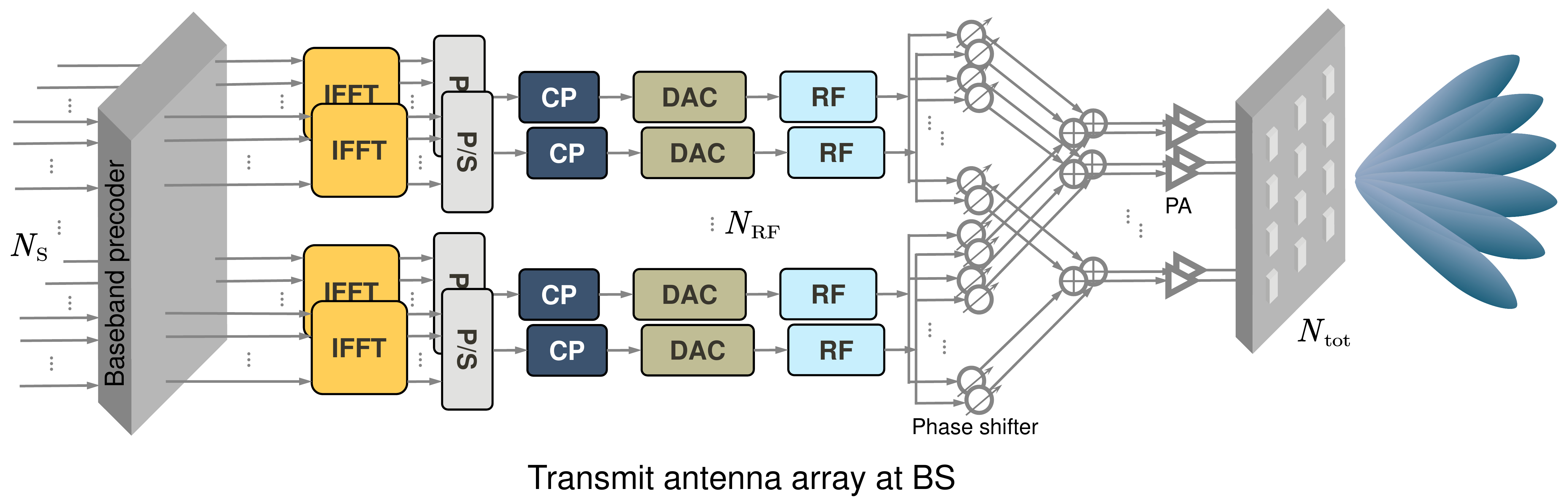}
\label{fig:subfigure1}}
\quad
\subfigure[]{%
\includegraphics[width=4.15in]{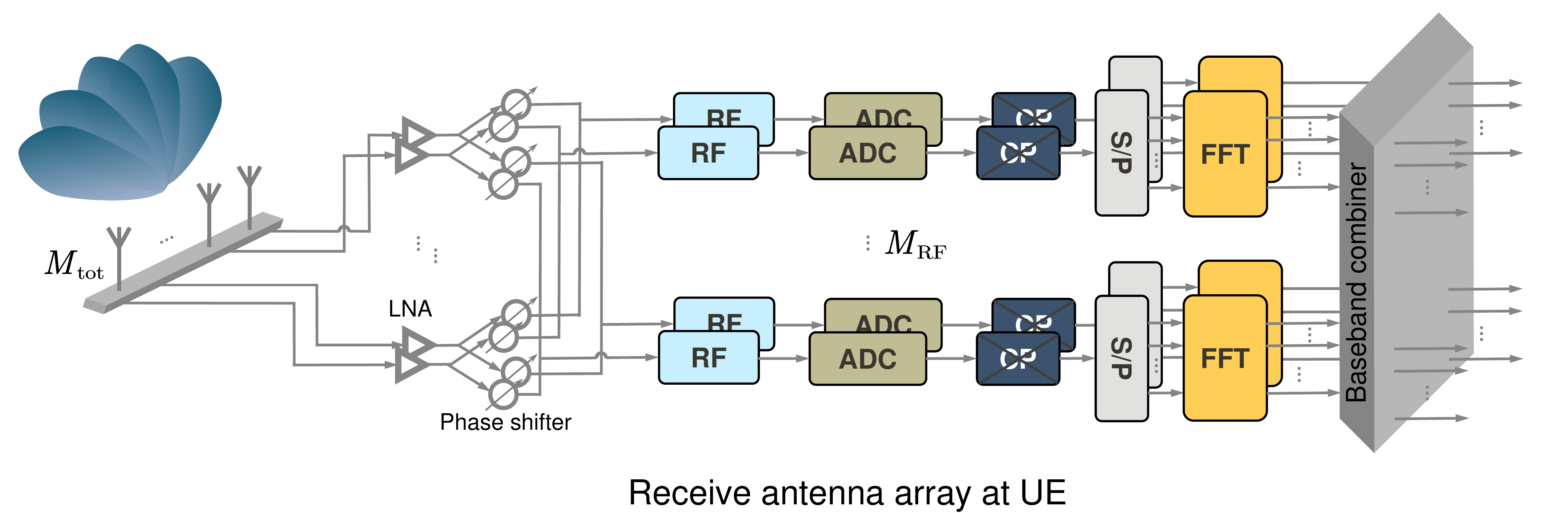}
\label{fig:subfigure2}}
\caption{(a) Shared-array architecture employed at the BS with $N_{\mathrm{RF}}$ RF chains and a total of $N_{\mathrm{tot}}$ transmit antenna elements configured in a uniform planar array. (b) Shared-array architecture employed at the UE with $M_{\mathrm{RF}}$ RF chains and a total of $M_{\mathrm{tot}}$ receive antenna elements configured in a uniform linear array.}
\label{fig:figure}
\end{figure}
In this section, we first present the employed system model including the transceiver architecture, antenna array configurations, and wideband mmWave channel model. We then illustrate the conventional grid-of-beams based beam tracking design along with an introduction to the frame structure.

\subsection{Transceiver architecture, antenna array configurations, and received signal model}
We consider a precoded MIMO-OFDM system with $N$ subcarriers and a hybrid precoding transceiver structure as shown in Figs.~1(a) and (b). A BS equipped with $N_{\mathrm{tot}}$ transmit antennas and $N_{\mathrm{RF}}$ RF chains transmits $N_{\mathrm{S}}$ data streams to a UE equipped with $M_{\mathrm{tot}}$ receive antennas and $M_{\mathrm{RF}}$ RF chains. As can be seen from Fig.~1, in a shared-array architecture, all antenna elements are jointly controlled by all RF chains sharing the same network of phase shifters. Further, we assume that a uniform planar array (UPA) is adopted at the BS, and a uniform linear array (ULA) is employed at the UE. The proposed methods are custom designed for uniform arrays, but can be extended to other array geometries by reconfiguring the beamforming vectors. The proposed methods are suited for both co-polarized and cross-polarized arrays \cite{dztrans2d}, though we focus on co-polarized array setup in this paper.

Based on the employed transceiver architecture, we now develop the baseband received signal model for our system after beamforming and combining. Let $\bm{s}[k]$ denote an $N_{\mathrm{S}}\times 1$ baseband transmit symbol vector such that $\mathbb{E}\left[\bm{s}[k]\bm{s}^{*}[k]\right]=\bm{I}_{N_{\mathrm{S}}}$ and $k=0,\cdots,N-1$. The data symbol vector $\bm{s}[k]$ is first precoded using an $N_{\mathrm{RF}}\times N_{\mathrm{S}}$ digital baseband precoding matrix $\bm{F}_{\mathrm{BB}}[k]$ on the $k$-th subcarrier, resulting in $\bm{d}[k]=\left[d_1[k],\cdots,d_{N_{\mathrm{RF}}}[k]\right]^{\mathrm{T}}=\bm{F}_{\mathrm{BB}}[k]\bm{s}[k]$. In this paper, we set $N_{\mathrm{RF}}=N_{\mathrm{S}}$ and $\bm{F}_{\mathrm{BB}}[k]=\bm{I}_{N_{\mathrm{S}}}$ because the channel tracking is conducted in the analog domain. Note that similar analog-only assumption applies to the UE side as well. The transmit symbols are then transformed to the time-domain via $N_{\mathrm{RF}}$, $N$-point IFFTs, generating the discrete-time signal sample $x_{n_{\mathrm{R}}}[n]=\sum_{k=0}^{N-1}d_{n_{\mathrm{R}}}[k]e^{\mathrm{j}\frac{2\pi k}{N}n}$, where $n_{\mathrm{R}}=1,\cdots,N_{\mathrm{RF}}$. Before applying an $N_{\mathrm{tot}}\times N_{\mathrm{RF}}$ wideband analog precoding matrix $\bm{F}_{\mathrm{RF}}$, a cyclic prefix (CP) with length $D$ is added to the data symbol blocks such that $D$ is greater than or equal to the maximum delay spread of the multi-path channel. Denote by $\bm{x}[n_{\mathrm{c}}]=\left[x_{1}[n_{\mathrm{c}}],\cdots,x_{N_{\mathrm{RF}}}[n_{\mathrm{c}}]\right]^{\mathrm{T}}$, where $n_{\mathrm{c}}=N-D,\cdots,N-1,0,\cdots,N-1$ due to the insertion of the CP. We can then express the discrete-time transmit signal model as $\bm{x}_{\mathrm{cp}}[n_{\mathrm{c}}] = \bm{F}_{\mathrm{RF}}\bm{x}[n_{\mathrm{c}}]$. To maintain the total transmit power constraint, $\left[\left[\bm{F}_{\mathrm{RF}}\right]_{:,n_{\mathrm{R}}}\left[\bm{F}_{\mathrm{RF}}\right]_{:,n_{\mathrm{R}}}^{*}\right]_{i,i}=\frac{1}{N_{\mathrm{tot}}}$ is satisfied with $i=1,\cdots,N_{\mathrm{tot}}$.

At the UE side, after combining with an $M_{\mathrm{tot}}\times M_{\mathrm{RF}}$ analog combining matrix $\bm{W}_{\mathrm{RF}}$, the CP is removed. The received data symbols are then transformed from the time-domain to the frequency-domain via $M_{\mathrm{RF}}$, $N$-point FFTs. Denote the frequency-domain $M_{\mathrm{tot}}\times N_{\mathrm{tot}}$ channel matrix by $\bm{H}[k]$. We can then express the discrete-time received signal as
\begin{equation}\label{resigsub}
\bm{y}[k]=\bm{W}_{\mathrm{RF}}^{*}\bm{H}[k]\bm{F}_{\mathrm{RF}}\bm{d}[k]+\bm{W}_{\mathrm{RF}}^{*}\bm{n}[k].
\end{equation}
The noise vector $\bm{n} \sim \mathcal{N}_{c}(\bm{0}_{M_{\mathrm{tot}}},\sigma^{2}\bm{I}_{M_{\mathrm{tot}}})$ and $\sigma^2=1/\gamma$, where $\gamma$ represents the target signal-to-noise ratio (SNR) before the transmit beamforming.

\subsection{Wideband channel model}
We employ a spatial geometric channel model to characterize the angular sparsity and frequency selectivity of the wideband mmWave channel. The spatial geometric channel models have been adopted in Long-Term Evolution (LTE) systems for various deployment scenarios \cite{ltech}. In Section V, we use practical channel parameters obtained via measurements to evaluate the proposed methods, though we employ the spatial geometric channel model to analytically explain the core idea. We assume that the channel has $N_{\mathrm{r}}$ paths, and each path $r$ has azimuth and elevation AoDs $\phi_{r}$, $\mu_{r}$, and AoA $\varphi_{r}$ in this paper. Let $p(\tau)$ denote the combined effect of filtering and pulse-shaping for $T_{\mathrm{s}}$-spaced signaling at $\tau$ seconds. We express the time-domain delay-$d$ MIMO channel matrix as
\begin{equation}\label{delayd}
\bm{H}[d]=\sum_{r=1}^{N_{\mathrm{r}}}g_{r}p\left(dT_{\mathrm{s}}-\tau_{r}\right)\bm{a}_{\mathrm{r}}(\varphi_{r})\bm{a}_{\mathrm{t}}^{*}(\mu_{r},\phi_{r}),
\end{equation}
where $g_{r}$ represents the complex path gain of path-$r$, $\bm{a}_{\mathrm{r}}(\cdot)\in\mathbb{C}^{M_{\mathrm{tot}}\times1}$ and $\bm{a}_{\mathrm{t}}(\cdot,\cdot)\in\mathbb{C}^{N_{\mathrm{tot}}\times1}$ correspond to the receive and transmit array response vectors. The channel frequency response matrix on subcarrier $k$ is the Fourier transform of $\bm{H}[d]$ such that
\begin{equation}\label{wbfrnp}
\bm{H}[k]=\sum_{r=1}^{N_{\mathrm{r}}}g_{r}\rho_{\tau_{r}}[k]\bm{a}_{\mathrm{r}}(\varphi_{r})\bm{a}_{\mathrm{t}}^{*}(\mu_{r},\phi_{r}),
\end{equation}
where $\rho_{\tau_{r}}[k]=\sum_{d=0}^{D-1}p\left(dT_{\mathrm{s}}-\tau_{r}\right)e^{-\mathrm{j}\frac{2\pi kd}{N}}$ is the Fourier transform of the delayed sampled filter $p(\tau)$ \cite{ahmedwb,wb2}.

Assuming that the UPA employed by the BS is in the $\mathrm{xy}$-plane with $N_{\mathrm{x}}$ and $N_{\mathrm{y}}$ elements on the $\mathrm{x}$ and $\mathrm{y}$ axes, then the transmit array response vector is
\begin{eqnarray}\label{upaco}
\bm{a}_{\mathrm{t}}(\mu_{r},\phi_{r})&=&\frac{1}{\sqrt{N_{\mathrm{tot}}}}\Big[1, e^{\mathrm{j}\frac{2\pi}{\lambda}d_{\mathrm{tx}}\sin(\mu_{r})\cos(\phi_{r})},\cdots,e^{\mathrm{j}\frac{2\pi}{\lambda}\left(N_{\mathrm{x}}-1\right)d_{\mathrm{tx}}\sin(\mu_{r})\cos(\phi_{r})}, e^{\mathrm{j}\frac{2\pi}{\lambda}d_{\mathrm{ty}}\sin(\mu_{r})\sin(\phi_{r})},\nonumber\\
&&\cdots,e^{\mathrm{j}\frac{2\pi}{\lambda}\left(\left(N_{\mathrm{x}}-1\right)d_{\mathrm{tx}}\sin(\mu_{r})\cos(\phi_{r})+\left(N_{\mathrm{y}}-1\right)d_{\mathrm{ty}}\sin(\mu_{r})\sin(\phi_{r})\right)}\Big]^{\mathrm{T}},
\end{eqnarray}
where $N_{\mathrm{tot}}=N_{\mathrm{x}}N_{\mathrm{y}}$, $\lambda$ represents the wavelength corresponding to the operating carrier frequency, $d_{\mathrm{tx}}$ and $d_{\mathrm{ty}}$ are the inter-element distances of the transmit antenna elements on the $\mathrm{x}$ and $\mathrm{y}$ axes. Denote by $\theta_r=\frac{2\pi}{\lambda}d_{\mathrm{tx}}\sin(\mu_r)\cos(\phi_r)$ and $\psi_r=\frac{2\pi}{\lambda}d_{\mathrm{ty}}\sin(\mu_r)\sin(\phi_r)$, which can be interpreted as the elevation and azimuth transmit spatial frequencies for path-$r$. We further define two vectors $\bm{a}_{\mathrm{tx}}(\theta_r)\in\mathbb{C}^{N_{\mathrm{x}}\times 1}$ and $\bm{a}_{\mathrm{ty}}(\psi_r)\in\mathbb{C}^{N_{\mathrm{y}}\times 1}$ as
\begin{eqnarray}
\bm{a}_{\mathrm{tx}}(\theta_r)=\frac{1}{\sqrt{N_{\mathrm{x}}}}\left[1, e^{\mathrm{j}\theta_r},\cdots, e^{\mathrm{j}\left(N_{\mathrm{x}}-1\right)\theta_r} \right]^{\mathrm{T}},\bm{a}_{\mathrm{ty}}(\psi_r)=\frac{1}{\sqrt{N_{\mathrm{y}}}}\left[1, e^{\mathrm{j}\psi_r},\cdots, e^{\mathrm{j}\left(N_{\mathrm{y}}-1\right)\psi_r} \right]^{\mathrm{T}},
\end{eqnarray}
which can be viewed as the transmit array response vectors in the elevation and azimuth domains. We therefore have $\bm{a}_{\mathrm{t}}(\theta_r,\psi_r)=\bm{a}_{\mathrm{tx}}(\theta_r)\otimes\bm{a}_{\mathrm{ty}}(\psi_r)$ \cite{dztrans2d}. With this decomposition, we are able to separately track the channel's azimuth and elevation angle information.

Since the ULA is employed by the UE, the receive array response vector is
\begin{equation}\label{ulaaaco}
\bm{a}_{\mathrm{r}}(\varphi_{r})=\frac{1}{\sqrt{M_{\mathrm{tot}}}}\big[1, e^{\mathrm{j}\frac{2\pi}{\lambda}d_{\mathrm{r}}\sin(\varphi_{r})},\cdots,e^{\mathrm{j}\frac{2\pi}{\lambda}d_{\mathrm{r}}\left(M_{\mathrm{tot}}-1\right)\sin(\varphi_{r})} \big]^{\mathrm{T}},
\end{equation}
where $d_{\mathrm{r}}$ denotes the inter-element distance between the receive antenna elements. Let $\nu_r=\frac{2\pi}{\lambda}d_{\mathrm{r}}\sin(\varphi_r)$ denote the receive spatial frequency for path-$r$. We can rewrite the receive array response vector for the UE as $\bm{a}_{\mathrm{r}}(\nu_r)=\frac{1}{\sqrt{M_{\mathrm{tot}}}}\left[1, e^{\mathrm{j}\nu_r},\cdots, e^{\mathrm{j}\left(M_{\mathrm{tot}}-1\right)\nu_r} \right]^{\mathrm{T}}$.

\begin{figure}
\centering
\subfigure[]{%
\includegraphics[width=2.35in]{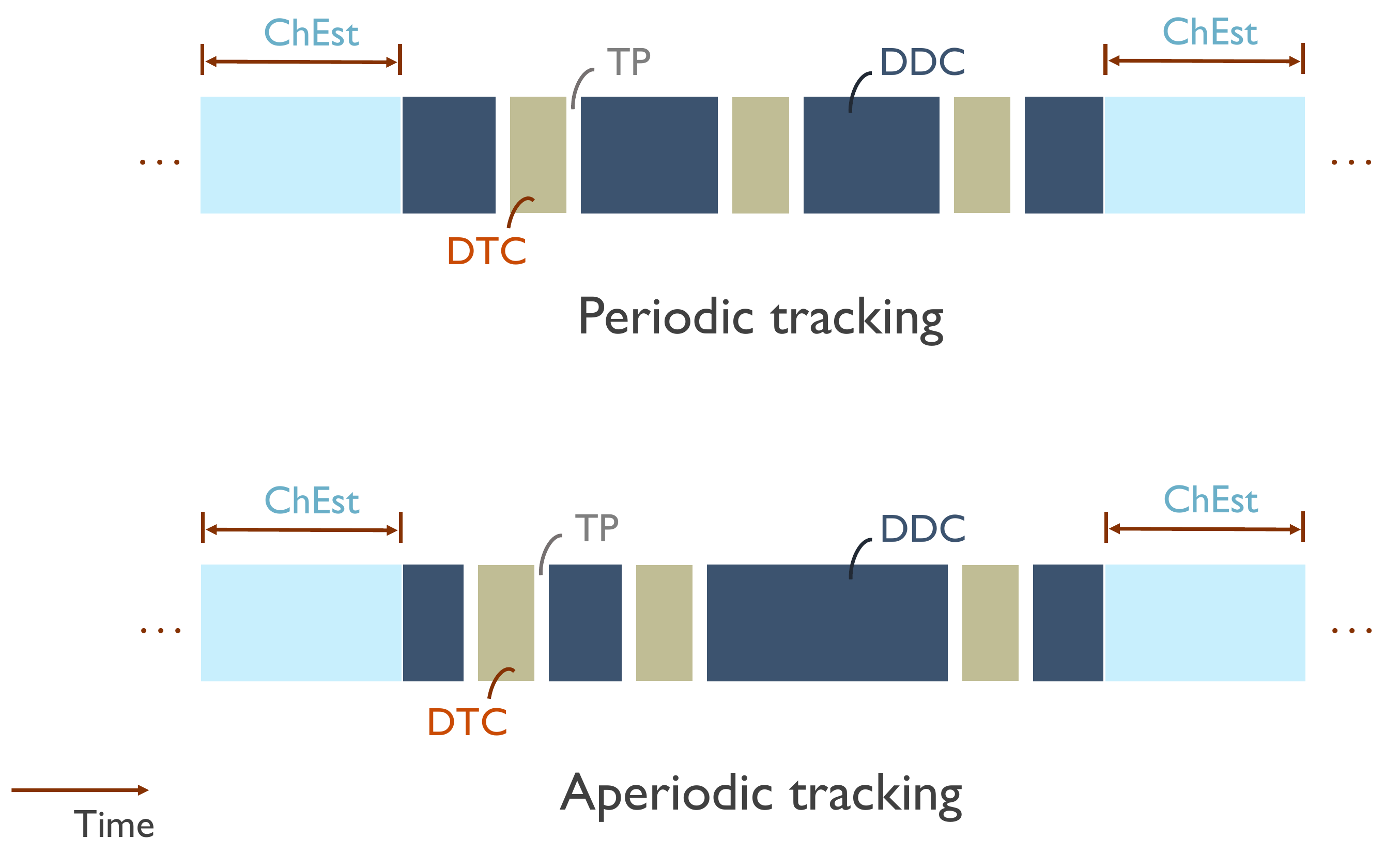}
\label{fig:subfigure1}}
\quad
\subfigure[]{%
\includegraphics[width=2.25in]{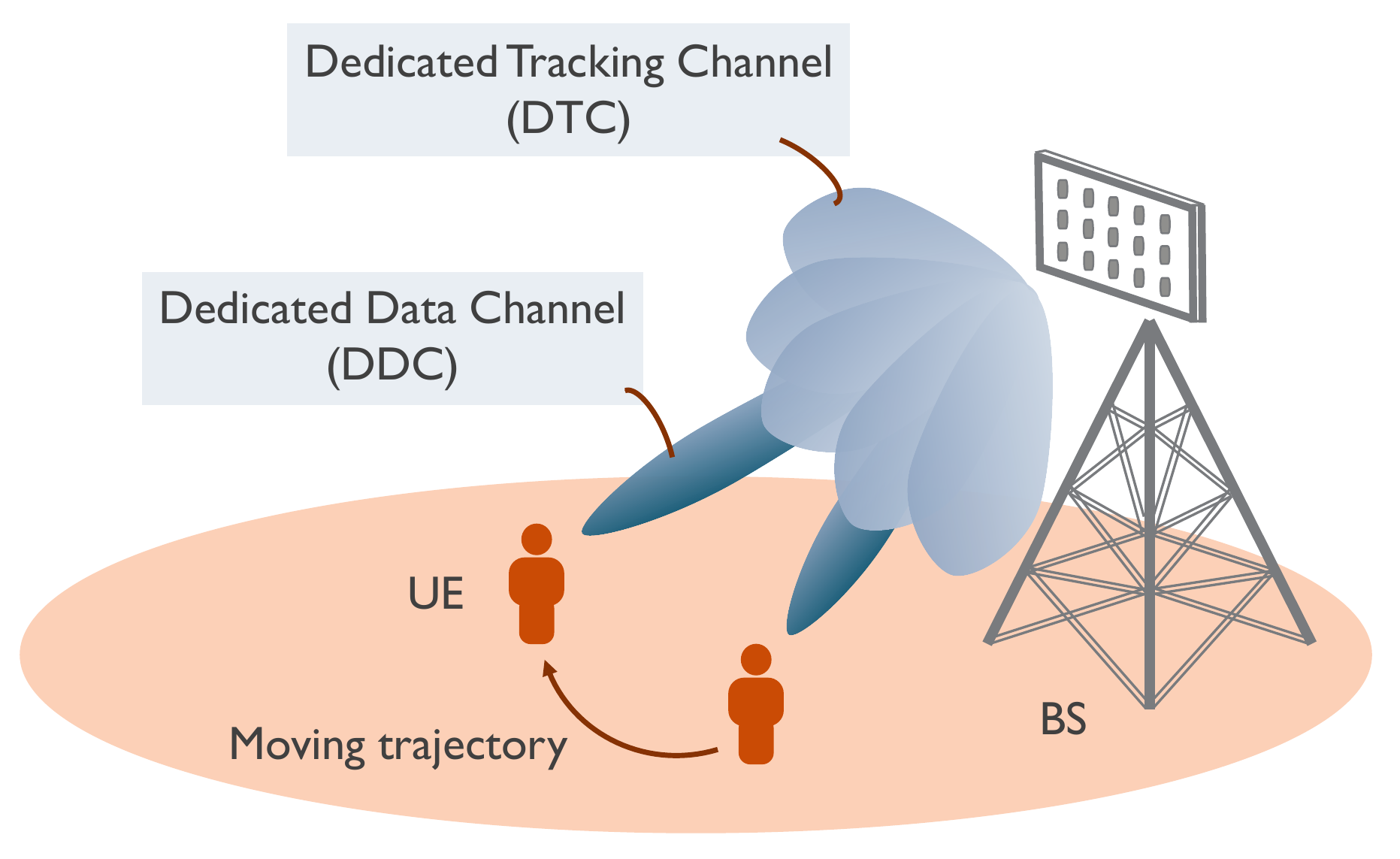}
\label{fig:subfigure2}}
\caption{(a) Potential frame structures of the periodic and aperidoic beam/angle tracking designs. For the periodic beam/angle tracking design, the periodicity of the dedicated tracking channel (DTC) is fixed. For the aperiodic beam/angle tracking design, the DTC is flexibly triggered and configured by the BS. The channel estimation (ChEst), dedicated data channel (DDC) and DTC are multiplexed in the time-domain. A transition period (TP) may exist between the DTC and the DDC. (b) One conceptual example of the multiplexing between the DDC and DTC. The steering directions of the beams in the DDC are adjusted towards the UE's positions, which are obtained via the tracking beams in the DTC.}
\label{fig:figure}
\end{figure}
\subsection{Frame structure and conventional beam tracking procedures}
In Fig.~2(a), we provide a potential frame structure. The time frame consists of three main components: channel estimation (ChEst), dedicated data channel (DDC), and dedicated tracking channel (DTC). The ChEst, DDC and DTC are composed of various numbers of time-slots. Here, we define the time-slot as the basic time unit, which is equivalent to, say, one OFDM symbol duration. We assume a total of $T$ time-slots for one DTC. In the DDC, directional narrow beams are probed by the BS for high-rate data communications, while in the DTC, relatively wide beams are used to track the channel variations. In this paper, the beams in the DTC and the DDC are multiplexed in the time-domain as shown in Fig.~2(a). A transition period (TP) may exist between the DTC and the DDC. Similar to the zero prefix/postfix design for OFDM \cite{prepostfix}, the TP is set as a zero region. As beams probed in the DDC and the DTC may have different beamwidths, the antenna array can be reconfigured during the TP. The TP may also handle the tracking requests and responses between the BS and the UE. Further, the beam tracking in the DTC can be conducted in either a periodic or an aperiodic manner as shown in Fig.~2(a). Based on the employed frame structure, we now illustrate the conventional grid-of-beams based beam tracking procedures for mmWave systems.

To reduce the computational complexity and tracking overhead, the beams in the DTC are formed surrounding the beam in the DDC in the angular domain. For simplicity, we first categorize all beams into three types, which are (i) the anchor beam (the beam in the DDC), (ii) the supporting beams (a predefined number of beams in the DTC that are closely surrounding the anchor beam), and (iii) the backup beams (the beams in the DTC other than the supporting beams). For a given DTC, the received signal strengths of the supporting and backup beams are measured by the UE and fed back to the BS, which are then compared with a predefined threshold. If the received signal strengths of the supporting beams are greater than the given threshold, the current anchor beam is continuously used for data communications until the next DTC is triggered. Otherwise, the backup beams that yield larger received signal strengths above the given threshold are considered, and the beam training is executed within the probing range of the selected backup beams to update the steering direction of the anchor beam. If the received signal strengths of all the supporting and backup beams are below the given threshold, the complete beam training process as in the channel estimation phase \cite{wang,singh2} will be conducted.

A conventional beam tracking design may incur a high tracking error probability due to the use of relatively wide beams and lack of quantization resolution \cite{bt4,btoverview}. Further, to update the steering direction of the anchor beam, an exhaustive search over all candidate anchor beams of interest is executed, which yields relatively high computational complexity and access delay. Hence, new beam or angle tracking methods with high tracking resolution and low implementation complexity are needed to enable reliable mmWave communications in fast-varying environments.

\section{Proposed Angle Tracking Designs for Mobile Wideband mmWave Systems}

In this section, we first illustrate the employed beam-specific pilot signal structure for the proposed tracking algorithms. Based on the employed shared-array architecture in Fig.~1, we then explain the design principles of the proposed high-resolution angle tracking approaches assuming the beam-specific pilot signal structure. Further, we present the detailed design procedures for the proposed algorithms along with the discussion of various feedback strategies. Unless otherwise specified, we explain the proposed angle tracking strategies in the azimuth domain assuming given elevation AoDs and AoAs. Note that the proposed algorithms can be directly extended to the tracking of the elevation directions.
\subsection{Design principles of proposed angle tracking approaches}
\begin{figure}
\centering
\subfigure[]{%
\includegraphics[width=2.35in]{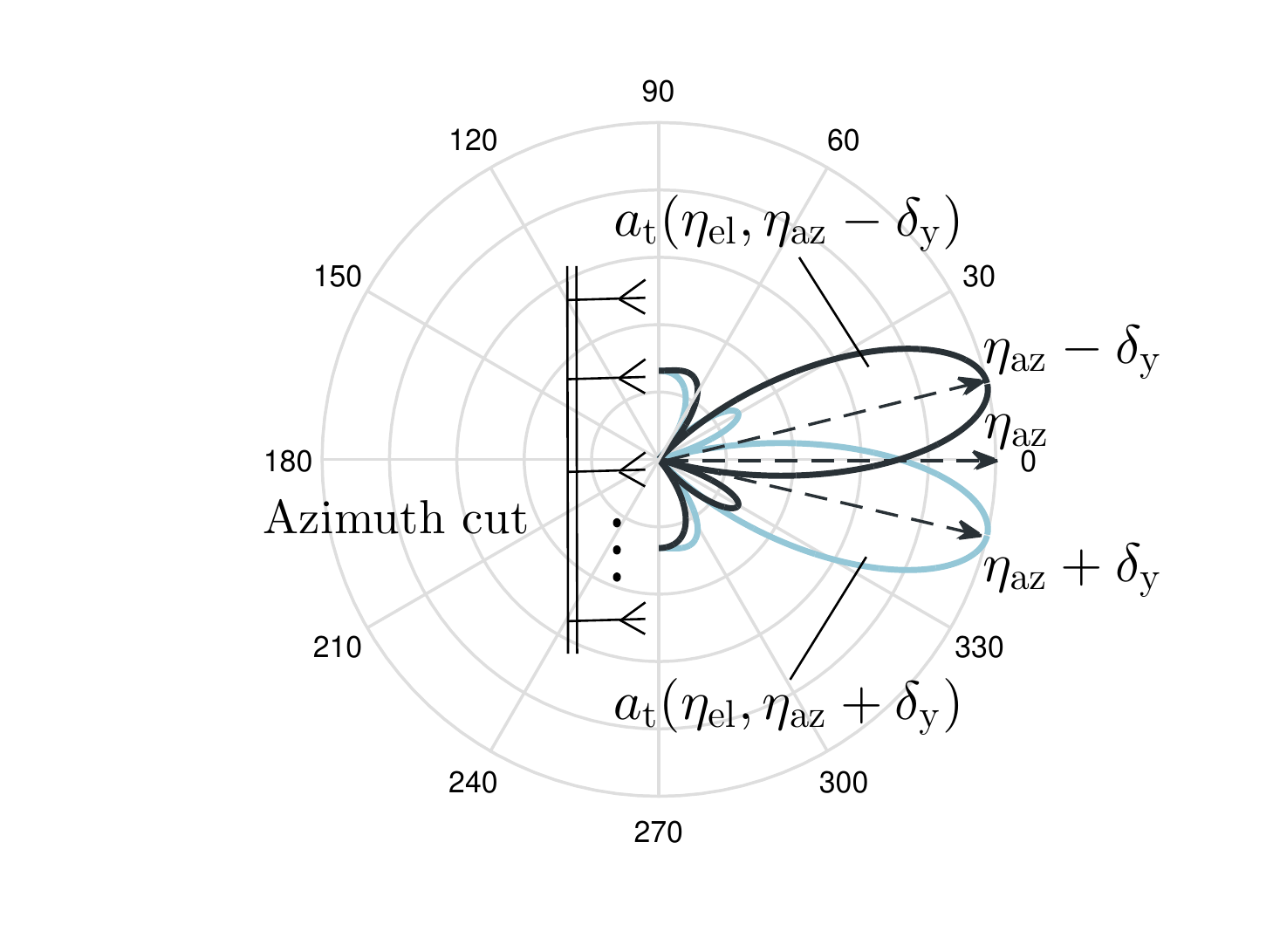}
\label{fig:subfigure1}}
\quad
\subfigure[]{%
\includegraphics[width=2.6in]{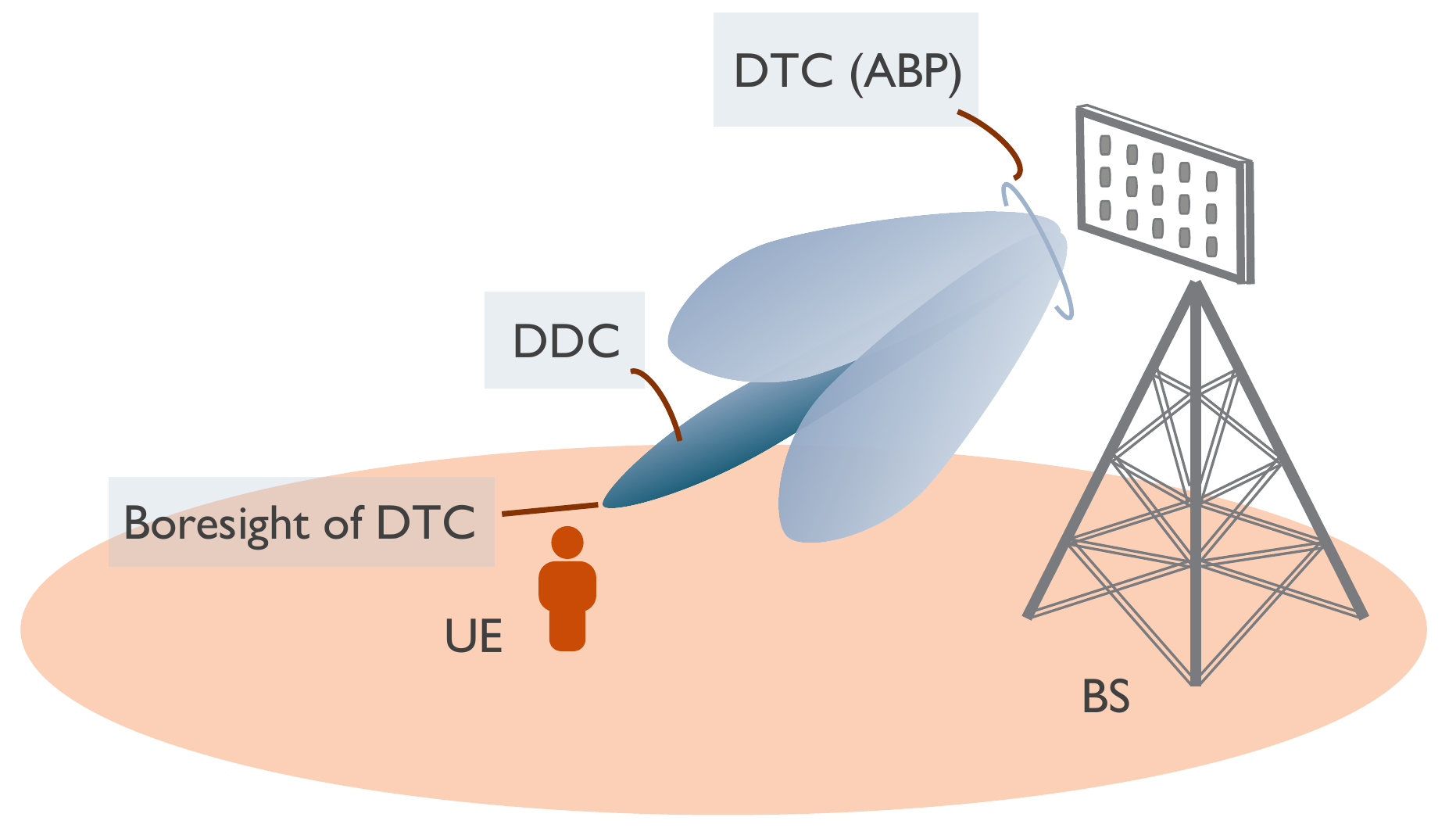}
\label{fig:subfigure2}}
\caption{(a) One conceptual example of the radiation patterns of one azimuth transmit auxiliary beam pair. The boresight angle of the corresponding auxiliary beam pair in the azimuth domain is $\eta_{\mathrm{az}}$, and the two beams steer towards $\eta_{\mathrm{az}}-\delta_{\mathrm{y}}$ and $\eta_{\mathrm{az}}+\delta_{\mathrm{y}}$ in the azimuth domain and $\eta_{\mathrm{el}}$ in the elevation domain. (b) One conceptual example of the relationship between the anchor beam in the DDC and the auxiliary beam pair (ABP) based tracking beams in the DTC. In this example, the steering direction of the anchor beam in the DDC is identical to the boresight angle of the auxiliary beam pair.}
\label{fig:figure}
\end{figure}
The design focus of the proposed angle tracking approaches is to first obtain high-resolution angle estimates, and then track the angle variations via the custom designed tracking beams. We employ the same frame structure as in Fig.~2(a) in the proposed methods, where the tracking beams are probed during the DTC. In this part, we provide an overview of the auxiliary beam pair-assisted high-resolution angle estimation design for wideband mmWave systems. For simplicity, we focus on the estimation of the azimuth AoDs at the receiver.

Each auxiliary beam pair comprises two successively probed analog beams in the angular domain. Pairs of custom designed analog transmit and receive beams are probed to cover the given angular ranges. In this paper, the two analog beams in the same auxiliary beam pair are formed simultaneously by the BS, and are differentiated by the beam-specific pilot signals at the UE side. In Fig.~3(a). we provide one conceptual example of the transmit auxiliary beam pair formed in the azimuth domain. As can be seen from Fig.~3(a), to form an azimuth transmit auxiliary beam pair, the two analog beamforming vectors targeted at the directions of $\eta_{\mathrm{az}}-\delta_{\mathrm{y}}$ and $\eta_{\mathrm{az}}+\delta_{\mathrm{y}}$ in the azimuth domain are $\bm{a}_{\mathrm{t}}(\eta_{\mathrm{el}},\eta_{\mathrm{az}}-\delta_{\mathrm{y}})$ and $\bm{a}_{\mathrm{t}}(\eta_{\mathrm{el}},\eta_{\mathrm{az}}+\delta_{\mathrm{y}})$, where $\delta_{\mathrm{y}}=\frac{2\ell_{\mathrm{y}}\pi}{N_{\mathrm{y}}}$ with $\ell_{\mathrm{y}}=1,\cdots,\frac{N_{\mathrm{y}}}{4}$ and $\eta_{\mathrm{el}}$ corresponds to a given elevation direction. Now, we illustrate the employed pilot signal structure.

Due to the constant amplitude and the robustness to the frequency selectivity, Zadoff-Chu (ZC)-type sequences are used in this paper as the pilot signals for tracking. Denoting the sequence length by $N_{\mathrm{ZC}}$, the employed ZC sequence with root index $i_z$ is
\begin{equation}\label{zcfdd}
s_{i_z}[m] = \exp\left(-\mathrm{j}\frac{\pi m(m+1)i_z}{N_{\mathrm{ZC}}}\right),
\end{equation}
where $m=0,\cdots,N_{\mathrm{ZC}}-1$. Here, we let $N_{\mathrm{ZC}}=N$ (i.e., the total number of employed subcarriers) and $z=0,1$ such that $i_0$ and $i_1$ correspond to the two beams in the same auxiliary beam pair. By cross correlating two ZC sequences at zero-lag, we can obtain \cite{popov}
\begin{equation}\label{fcorrex}
\sum_{k=0}^{N-1}s_{i_0}[k]s^{*}_{i_1}[k]=\Bigg\{\begin{array}{l}
                                                    1,\hspace{2mm} \textrm{if}\hspace{2mm} i_0=i_1 \\
                                                    \beta_{i_0,i_1},\hspace{2mm}\textrm{otherwise}.
                                                  \end{array}
\end{equation}
Here, $\beta_{i_0,i_1}$ is a constant with small magnitude $\left|\beta_{i_0,i_1}\right|\approx 0$. In this paper, we assume $\beta_{i_0,i_1}=0$, i.e., the two sequences are orthogonal in the code-domain. By leveraging this code-domain orthogonality, the two simultaneously probed beams in the same auxiliary beam pair can be differentiated by the UE without interference.

Based on the employed pilot signal structure, we now explain the design principles of the auxiliary beam pair-assisted angle acquisition. Assume $M_{\mathrm{RF}}=1$ and a given analog receive beam, say, $\bm{a}_{\mathrm{r}}(\vartheta)$. According to the employed array configurations and the pilot signal structure, we can then rewrite (\ref{resigsub}) in the absence of noise as
\begin{equation}\label{resigsubnew}
y[k]=\bm{a}^{*}_{\mathrm{r}}(\vartheta)\sum_{r=1}^{N_{\mathrm{r}}}g_{r}\rho_{\tau_{r}}[k]\bm{a}_{\mathrm{r}}(\nu_{r})\bm{a}_{\mathrm{t}}^{*}(\theta_{r},\psi_{r})\left[
                                                  \begin{array}{cc}
                                                    \bm{a}_{\mathrm{t}}(\eta_{\mathrm{el}},\eta_{\mathrm{az}}-\delta_{\mathrm{y}}) & \bm{a}_{\mathrm{t}}(\eta_{\mathrm{el}},\eta_{\mathrm{az}}+\delta_{\mathrm{y}}) \\
                                                  \end{array}
                                                \right]
\left[
  \begin{array}{c}
    s_{i_0}[k] \\
    s_{i_1}[k] \\
  \end{array}
\right].
\end{equation}
Our design focus here is to estimate the azimuth transmit spatial frequency $\psi_{r^{\star}}$ for path-$r^{\star}$ with $r^{\star}\in\left\{1,\cdots,N_{\mathrm{r}}\right\}$. We first assume that $\psi_{r^{\star}}$ falls into the probing range of the auxiliary beam pair such that $\psi_{r^{\star}}\in\left(\eta_{\mathrm{az}}-\delta_{\mathrm{y}},\eta_{\mathrm{az}}+\delta_{\mathrm{y}}\right)$. This is possible by first selecting the $N_{\mathrm{r}}$ beams with the largest received powers, and then $N_{\mathrm{r}}$ auxiliary beam pairs according to \cite[Lemma~2]{dztrans} such that each spatial frequency can be covered by the corresponding auxiliary beam pair with high probability. We can then rewrite (\ref{resigsubnew}) as
\begin{align}\label{resigsubnewsep}
y[k]&=\bm{a}^{*}_{\mathrm{r}}(\vartheta)g_{r^{\star}}\rho_{\tau_{r^{\star}}}[k]\bm{a}_{\mathrm{r}}(\nu_{r^{\star}})\bm{a}_{\mathrm{t}}^{*}(\theta_{r^{\star}},\psi_{r^{\star}})\left[
                                                  \begin{array}{cc}
                                                    \bm{a}_{\mathrm{t}}(\eta_{\mathrm{el}},\eta_{\mathrm{az}}-\delta_{\mathrm{y}}) & \bm{a}_{\mathrm{t}}(\eta_{\mathrm{el}},\eta_{\mathrm{az}}+\delta_{\mathrm{y}}) \\
                                                  \end{array}
                                                \right]
\left[
  \begin{array}{c}
    s_{i_0}[k] \\
    s_{i_1}[k] \\
  \end{array}
\right]\nonumber\\
&+\underbrace{\bm{a}^{*}_{\mathrm{r}}(\vartheta)\sum_{\substack{r'=1,\\r'\neq r^{\star}}}^{N_{\mathrm{r}}}g_{r'}\rho_{\tau_{r'}}[k]\bm{a}_{\mathrm{r}}(\nu_{r'})\bm{a}_{\mathrm{t}}^{*}(\theta_{r'},\psi_{r'})\left[
                                                  \begin{array}{cc}
                                                    \bm{a}_{\mathrm{t}}(\eta_{\mathrm{el}},\eta_{\mathrm{az}}-\delta_{\mathrm{y}}) & \bm{a}_{\mathrm{t}}(\eta_{\mathrm{el}},\eta_{\mathrm{az}}+\delta_{\mathrm{y}}) \\
                                                  \end{array}
                                                \right]
\left[
  \begin{array}{c}
    s_{i_0}[k] \\
    s_{i_1}[k] \\
  \end{array}
\right]}_{\textrm{multi-path interference}}.
\end{align}
Because of the angular sparsity of the mmWave channels \cite{dztrans2d}, we assume that other paths' spatial frequencies are not covered by the auxiliary beam pair, i.e., $\psi_{r'}\notin\left(\eta_{\mathrm{az}}-\delta_{\mathrm{y}},\eta_{\mathrm{az}}+\delta_{\mathrm{y}}\right)$ with $r'\in\left\{1,\cdots,N_{\mathrm{r}}\right\}$ and $r'\neq r^{\star}$. Along with the assumption of the high-power regime (e.g., $N_{\mathrm{y}}\rightarrow\infty$), we ignore the multi-path interference and rewrite (\ref{resigsubnewsep}) as
\begin{equation}\label{resigsubnewag}
y[k]=g_{r^{\star}}\rho_{\tau_{r^{\star}}}[k]\bm{a}^{*}_{\mathrm{r}}(\vartheta)\bm{a}_{\mathrm{r}}(\nu_{r^{\star}})\bm{a}_{\mathrm{t}}^{*}(\theta_{r^{\star}},\psi_{r^{\star}})\left[
                                                  \begin{array}{cc}
                                                    \bm{a}_{\mathrm{t}}(\eta_{\mathrm{el}},\eta_{\mathrm{az}}-\delta_{\mathrm{y}}) & \bm{a}_{\mathrm{t}}(\eta_{\mathrm{el}},\eta_{\mathrm{az}}+\delta_{\mathrm{y}}) \\
                                                  \end{array}
                                                \right]
\left[
  \begin{array}{c}
    s_{i_0}[k] \\
    s_{i_1}[k] \\
  \end{array}
\right].
\end{equation}
Note that we can extend the algorithm to separately estimate multiple paths in parallel.

Assuming perfect time-frequency synchronization, the UE employs locally stored reference beam-specific sequences to correlate the received signal samples. By using the reference ZC sequence with the sequence root index $i_0$, we can first obtain
\begin{eqnarray}
\Lambda^{\Delta}_{\mathrm{az}}&=&\sum_{k=0}^{N-1}s^{*}_{i_0}[k]y[k]\\
&=&\sum_{k=0}^{N-1}g_{r^{\star}}\rho_{\tau_{r^{\star}}}[k]\bm{a}^{*}_{\mathrm{r}}(\vartheta)\bm{a}_{\mathrm{r}}(\nu_{r^{\star}})\bm{a}_{\mathrm{t}}^{*}(\theta_{r^{\star}},\psi_{r^{\star}})\bm{a}_{\mathrm{t}}(\eta_{\mathrm{el}},\eta_{\mathrm{az}}-\delta_{\mathrm{y}})s^{*}_{i_0}[k]s_{i_0}[k]\nonumber\\
&+&\sum_{k=0}^{N-1}g_{r^{\star}}\rho_{\tau_{r^{\star}}}[k]\bm{a}^{*}_{\mathrm{r}}(\vartheta)\bm{a}_{\mathrm{r}}(\nu_{r^{\star}})\bm{a}_{\mathrm{t}}^{*}(\theta_{r^{\star}},\psi_{r^{\star}})\bm{a}_{\mathrm{t}}(\eta_{\mathrm{el}},\eta_{\mathrm{az}}+\delta_{\mathrm{y}})s^{*}_{i_0}[k]s_{i_1}[k].\label{flatccch}
\end{eqnarray}
We assume flat channels here such that $\bar{\rho}_{\tau_{r^{\star}}}=\rho_{\tau_{r^{\star}}}[0]=\cdots=\rho_{\tau_{r^{\star}}}[N-1]$ for better illustration of the design principles. The proposed design approach can still achieve promising angle estimation/tracking performance in wideband channels (verified in Section V-B) since the correlation properties of the ZC-type sequences are robust to the frequency selectivity (e.g., up to $8.6$ MHz continuous bandwidth in LTE \cite{lte, zcfreq}). We can then rewrite (\ref{flatccch}) as
\begin{eqnarray}
\Lambda^{\Delta}_{\mathrm{az}}&=&g_{r^{\star}}\bm{a}^{*}_{\mathrm{r}}(\vartheta)\bm{a}_{\mathrm{r}}(\nu_{r^{\star}})\bm{a}_{\mathrm{t}}^{*}(\theta_{r^{\star}},\psi_{r^{\star}})\bm{a}_{\mathrm{t}}(\eta_{\mathrm{el}},\eta_{\mathrm{az}}-\delta_{\mathrm{y}})\bar{\rho}_{\tau_{r^{\star}}}\sum_{k=0}^{N-1}s^{*}_{i_0}[k]s_{i_0}[k]\nonumber\\
&+&g_{r^{\star}}\bm{a}^{*}_{\mathrm{r}}(\vartheta)\bm{a}_{\mathrm{r}}(\nu_{r^{\star}})\bm{a}_{\mathrm{t}}^{*}(\theta_{r^{\star}},\psi_{r^{\star}})\bm{a}_{\mathrm{t}}(\eta_{\mathrm{el}},\eta_{\mathrm{az}}+\delta_{\mathrm{y}})\bar{\rho}_{\tau_{r^{\star}}}\sum_{k=0}^{N-1}s^{*}_{i_0}[k]s_{i_1}[k]\\
&\overset{(a)}{=}&g_{r^{\star}}\bar{\rho}_{\tau_{r^{\star}}}\bm{a}^{*}_{\mathrm{r}}(\vartheta)\bm{a}_{\mathrm{r}}(\nu_{r^{\star}})\bm{a}_{\mathrm{t}}^{*}(\theta_{r^{\star}},\psi_{r^{\star}})\bm{a}_{\mathrm{t}}(\eta_{\mathrm{el}},\eta_{\mathrm{az}}-\delta_{\mathrm{y}}),
\end{eqnarray}
where ($a$) is due to the employed beam-specific pilot signal structure in (\ref{fcorrex}). We then compute the corresponding received signal strength as
\begin{eqnarray}\label{neglect}
\chi^{\Delta}_{\mathrm{az}}&=&\left(\Lambda^{\Delta}_{\mathrm{az}}\right)^{*}\Lambda^{\Delta}_{\mathrm{az}}\\
&=&\left|g_{r^{\star}}\bar{\rho}_{\tau_{r^{\star}}}\right|^{2}\left|\bm{a}^{*}_{\mathrm{r}}(\vartheta)\bm{a}_{\mathrm{r}}(\nu_{r^{\star}})\right|^{2}\nonumber\\
&\times&\bm{a}^{*}_{\mathrm{t}}(\eta_{\mathrm{el}},\eta_{\mathrm{az}}-\delta_{\mathrm{y}})\bm{a}_{\mathrm{t}}(\theta_{r^{\star}},\psi_{r^{\star}})\bm{a}_{\mathrm{t}}^{*}(\theta_{r^{\star}},\psi_{r^{\star}})\bm{a}_{\mathrm{t}}(\eta_{\mathrm{el}},\eta_{\mathrm{az}}-\delta_{\mathrm{y}}).
\end{eqnarray}
Similarly, using the ZC sequence with the root index $i_1$ to correlate the received signal samples, we obtain
\begin{eqnarray}
\Lambda^{\Sigma}_{\mathrm{az}}=\sum_{k=0}^{N-1}s^{*}_{i_1}[k]y[k]=g_{r^{\star}}\bar{\rho}_{\tau_{r^{\star}}}\bm{a}^{*}_{\mathrm{r}}(\vartheta)\bm{a}_{\mathrm{r}}(\nu_{r^{\star}})\bm{a}_{\mathrm{t}}^{*}(\theta_{r^{\star}},\psi_{r^{\star}})\bm{a}_{\mathrm{t}}(\eta_{\mathrm{el}},\eta_{\mathrm{az}}+\delta_{\mathrm{y}}).
\end{eqnarray}
We can calculate the corresponding received signal strength as
\begin{eqnarray}\label{neglectag}
\chi^{\Sigma}_{\mathrm{az}}&=&\left(\Lambda^{\Sigma}_{\mathrm{az}}\right)^{*}\Lambda^{\Sigma}_{\mathrm{az}}\\
&=&\left|g_{r^{\star}}\bar{\rho}_{\tau_{r^{\star}}}\right|^{2}\left|\bm{a}^{*}_{\mathrm{r}}(\vartheta)\bm{a}_{\mathrm{r}}(\nu_{r^{\star}})\right|^{2}\nonumber\\
&\times&\bm{a}^{*}_{\mathrm{t}}(\eta_{\mathrm{el}},\eta_{\mathrm{az}}+\delta_{\mathrm{y}})\bm{a}_{\mathrm{t}}(\theta_{r^{\star}},\psi_{r^{\star}})\bm{a}_{\mathrm{t}}^{*}(\theta_{r^{\star}},\psi_{r^{\star}})\bm{a}_{\mathrm{t}}(\eta_{\mathrm{el}},\eta_{\mathrm{az}}+\delta_{\mathrm{y}}).
\end{eqnarray}
We can further express $\chi^{\Delta}_{\mathrm{az}}$ and $\chi^{\Sigma}_{\mathrm{az}}$ as
\begin{eqnarray}
\chi^{\Delta}_{\mathrm{az}}&=& \left|g_{r^{\star}}\bar{\rho}_{\tau_{r^{\star}}}\right|^{2}\left|\bm{a}^{*}_{\mathrm{r}}(\vartheta)\bm{a}_{\mathrm{r}}(\nu_{r^{\star}})\right|^{2}\frac{\sin^{2}\left(\frac{N_{\mathrm{x}}(\theta_{r^{\star}}-\eta_{\mathrm{el}})}{2}\right)}{\sin^{2}\left(\frac{\theta_{r^{\star}}-\eta_{\mathrm{el}}}{2}\right)}\frac{\sin^{2}\left(\frac{N_{\mathrm{y}}(\psi_{r^{\star}}-\eta_{\mathrm{az}})}{2}\right)}{\sin^{2}\left(\frac{\psi_{r^{\star}}-\eta_{\mathrm{az}}+\delta_{\mathrm{y}}}{2}\right)},\\
\chi^{\Sigma}_{\mathrm{az}}&=& \left|g_{r^{\star}}\bar{\rho}_{\tau_{r^{\star}}}\right|^{2}\left|\bm{a}^{*}_{\mathrm{r}}(\vartheta)\bm{a}_{\mathrm{r}}(\nu_{r^{\star}})\right|^{2}\frac{\sin^{2}\left(\frac{N_{\mathrm{x}}(\theta_{r^{\star}}-\eta_{\mathrm{el}})}{2}\right)}{\sin^{2}\left(\frac{\theta_{r^{\star}}-\eta_{\mathrm{el}}}{2}\right)}\frac{\sin^{2}\left(\frac{N_{\mathrm{y}}(\psi_{r^{\star}}-\eta_{\mathrm{az}})}{2}\right)}{\sin^{2}\left(\frac{\psi_{r^{\star}}-\eta_{\mathrm{az}}-\delta_{\mathrm{y}}}{2}\right)},\label{expaaaa}
\end{eqnarray}
where (21) and (\ref{expaaaa}) are obtained via $\left|\sum_{\bar{m}=1}^{M}e^{-\mathrm{j}(\bar{m}-1)\bar{x}}\right|^{2}=\frac{\sin^{2}\left(\frac{M\bar{x}}{2}\right)}{\sin^{2}\left(\frac{\bar{x}}{2}\right)}$.
We define the ratio metric
\begin{eqnarray}\label{rmetric}
\zeta_{\mathrm{az}}=\frac{\chi^{\Delta}_{\mathrm{az}}-\chi^{\Sigma}_{\mathrm{az}}}{\chi^{\Delta}_{\mathrm{az}}+\chi^{\Sigma}_{\mathrm{az}}}=\frac{\sin^{2}\left(\frac{\psi_{r^{\star}}-\eta_{\mathrm{az}}-\delta_{\mathrm{y}}}{2}\right)-\sin^{2}\left(\frac{\psi_{r^{\star}}-\eta_{\mathrm{az}}+\delta_{\mathrm{y}}}{2}\right)}{\sin^{2}\left(\frac{\psi_{r^{\star}}-\eta_{\mathrm{az}}-\delta_{\mathrm{y}}}{2}\right)+\sin^{2}\left(\frac{\psi_{r^{\star}}-\eta_{\mathrm{az}}+\delta_{\mathrm{y}}}{2}\right)}=-\frac{\sin\left(\psi_{r^{\star}}-\eta_{\mathrm{az}}\right)\sin(\delta_{\mathrm{y}})}{1-\cos\left(\psi_{r^{\star}}-\eta_{\mathrm{az}}\right)\cos(\delta_{\mathrm{y}})}.
\end{eqnarray}
According to \cite[Lemma~1]{dztrans}, if $|\psi_{r^{\star}}-\eta_{\mathrm{az}}|<\delta_{\mathrm{y}}$, then the azimuth transmit spatial frequency $\psi_{r^{\star}}$ is within the range of $\left(\eta_{\mathrm{az}}-\delta_{\mathrm{y}},\eta_{\mathrm{az}}+\delta_{\mathrm{y}}\right)$, and $\zeta_{\mathrm{az}}$ is a monotonically decreasing function of $\psi_{r^{\star}}-\eta_{\mathrm{az}}$ and invertible with respect to $\psi_{r^{\star}}-\eta_{\mathrm{az}}$. Via the inverse function, we can therefore derive the estimated value of $\psi_{r^{\star}}$ as
\begin{equation}\label{angleestimate}
\hat{\psi}_{r^{\star}}=\eta_{\mathrm{az}}-\arcsin\left(\frac{\zeta_{\mathrm{az}}\sin(\delta_{\mathrm{y}})-\zeta_{\mathrm{az}}\sqrt{1-\zeta_{\mathrm{az}}^{2}}\sin(\delta_{\mathrm{y}})\cos(\delta_{\mathrm{y}})}{\sin^{2}(\delta_{\mathrm{y}})+\zeta_{\mathrm{az}}^{2}\cos^{2}(\delta_{\mathrm{y}})}\right).
\end{equation}
If $\zeta_{\mathrm{az}}$ is perfect, i.e., not impaired by noise and other types of interference, we can perfectly recover the azimuth transmit spatial frequency for path-$r^{\star}$, i.e., $\psi_{r^{\star}}=\hat{\psi}_{r^{\star}}$.

In Section III-B, we restrict to the tracking of path-$r^{\star}$'s azimuth AoD. To better reveal the temporal evolution, we use $\psi_{r^{\star},t}$ instead of $\psi_{r^{\star}}$ to represent path-$r^{\star}$'s azimuth transmit spatial frequency for a given time-slot $t\in\left\{0,\cdots,T-1\right\}$ in the DTC.
\subsection{Design procedures of proposed angle tracking approaches}

Leveraging the high-resolution angle estimates, we exploit the auxiliary beam pair design in forming tracking beams in the DTC. In Fig.~3(b), we present one conceptual example of applying the auxiliary beam pair based approach in angle tracking. In this example, one transmit auxiliary beam pair (e.g., $\bm{a}_{\mathrm{t}}(\eta_{\mathrm{el}},\eta_{\mathrm{az}}-\delta_{\mathrm{y}})$ and $\bm{a}_{\mathrm{t}}(\eta_{\mathrm{el}},\eta_{\mathrm{az}}+\delta_{\mathrm{y}})$ in Fig.~3(a)) is probed during the DTC. The boresight angle of the auxiliary beam pair (e.g., $\eta_{\mathrm{az}}$ in Fig.~3(a)) is identical to the steering direction of the corresponding anchor beam in the DDC. In the following, we first illustrate the general framework of the proposed angle tracking designs.

\begin{figure}
\centering
\includegraphics[width=5.75in]{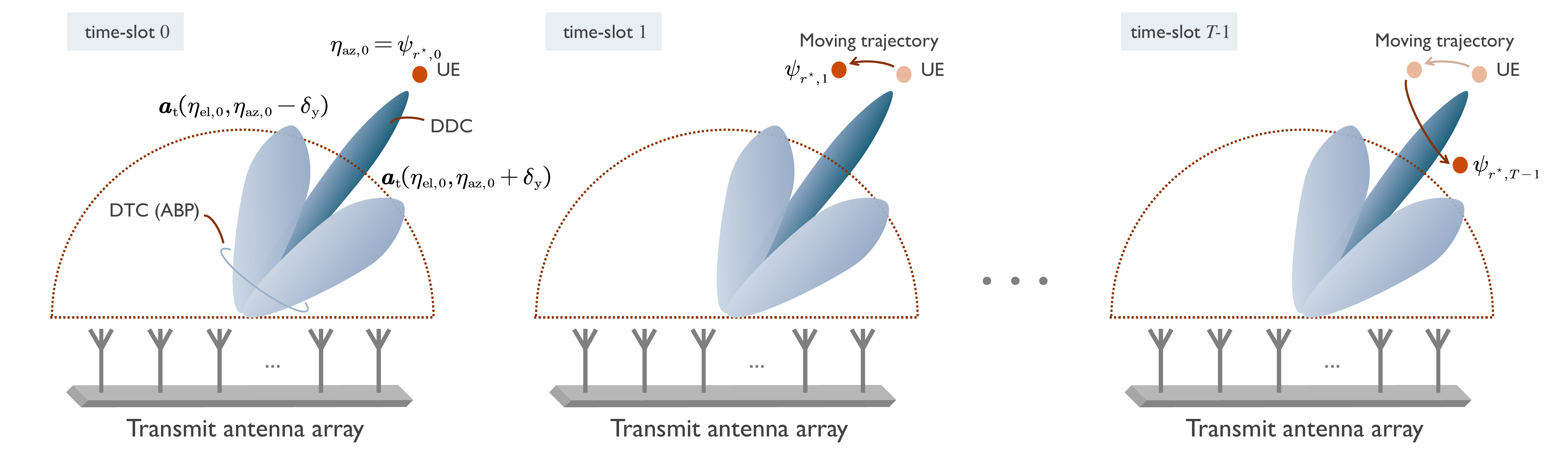}
\label{fig:subfigure1}
\caption{Conceptual examples of the relationship between the UE's moving trajectory (i.e., the angle variations) and the auxiliary beam pair based tracking beams in the DTC. As long as the relative position of the UE to the transmit antenna array is within the probing range of the auxiliary beam pair, it is expected to be tracked via the tracking beams in the DTC.}
\label{fig:figure}
\end{figure}
In Fig.~4, we provide the relationship between the UE's moving trajectory and the tracking beams in the DTC. At time-slot $0$, the anchor beam in the DDC with the azimuth boresight angle $\eta_{\mathrm{az},0}$ steers towards the UE of interest. One azimuth transmit auxiliary beam pair is formed as the tracking beams in the DTC. For a given elevation direction $\eta_{\mathrm{el},0}$, the corresponding two beams probe towards $\eta_{\mathrm{az},0}-\delta_{\mathrm{y}}$ and $\eta_{\mathrm{az},0}+\delta_{\mathrm{y}}$ with the boresight angle $\eta_{\mathrm{az},0}$ in the azimuth domain. As can be seen from the conceptual example shown in Fig.~4, at time-slots $1,\cdots,T-1$, the UE of interest moves away from the original azimuth position $\psi_{r^{\star},0}$ (or $\eta_{\mathrm{az},0}$) to $\psi_{r^{\star},1},\cdots,\psi_{r^{\star},T-1}$. Note that as long as $\psi_{r^{\star},1},\cdots,\psi_{r^{\star},T-1}$ are in the probing range of the tracking beams, they are expected to be accurately tracked according to the design principles of the auxiliary beam pair.

In the proposed methods, either the BS or the UE can trigger the angle tracking process, which are referred to as BS-driven or UE-driven angle tracking methods. For both the BS-driven and UE-driven angle tracking strategies, either a periodic or aperiodic DTC design can be adopted. Further, for the proposed BS-driven angle tracking, no prior knowledge of the auxiliary beam pair setup is required at the UE side. In the following, we present the detailed design procedures of the proposed methods and illustrate the employed direct and differential feedback strategies.

\emph{1. BS/UE-driven angle tracking design with direct ratio metric feedback}. We start by illustrating the BS-driven angle tracking strategy using the direct ratio metric feedback. For a given time-slot $t\in\left\{0,\cdots,T-1\right\}$ in the DTC, the corresponding ratio metric $\zeta_{\mathrm{az},t}$ is calculated by the UE according to (\ref{rmetric}) using the probed azimuth transmit auxiliary beam pair. First, assume that the BS triggers the feedback of the derived ratio metric. For instance, considering a given DTC, if the BS requires the ratio metric feedback at time-slot $T-1$, $\zeta_{\mathrm{az},T-1}$ is then quantized and sent back to the BS. In this case, time-slot $T-1$ is the last time-slot of a given DTC. Note that in practice, the BS may require the ratio metric feedback for multiple time-slots within the same DTC to track the fast-varying channels. It is therefore essential for the UE to keep computing the ratio metric for every time-slot in the DTC. Upon receiving the ratio metric feedback from the UE at time-slot $t$, the BS retrieves the corresponding angle estimate according to (\ref{angleestimate}). Denoting the azimuth angle estimate at time-slot $t$ by $\hat{\psi}_{r^{\star},t}$, we have
\begin{equation}\label{newangleestimate}
\hat{\psi}_{r^{\star},t}=\eta_{\mathrm{az},0}-\arcsin\left(\frac{\zeta_{\mathrm{az},t}\sin(\delta_{\mathrm{y}})-\zeta_{\mathrm{az},t}\sqrt{1-\zeta_{\mathrm{az},t}^{2}}\sin(\delta_{\mathrm{y}})\cos(\delta_{\mathrm{y}})}{\sin^{2}(\delta_{\mathrm{y}})+\zeta_{\mathrm{az},t}^{2}\cos^{2}(\delta_{\mathrm{y}})}\right).
\end{equation}
The angle difference $\Delta\psi_{r^{\star},t}=\left|\psi_{r^{\star},0}-\hat{\psi}_{r^{\star},t}\right|$ is then calculated by the BS and compared with a predefined threshold $\varsigma_{\mathrm{az}}$. If $\Delta\psi_{r^{\star},t}\geq\varsigma_{\mathrm{az}}$, the azimuth steering direction of the anchor beam in the DDC is then updated from $\eta_{\mathrm{az},0}$ to $\eta_{\mathrm{az},t}=\hat{\psi}_{r^{\star},t}$. Otherwise, the azimuth steering direction of the anchor beam in the DDC is kept unchanged from time-slot $0$, i.e., $\eta_{\mathrm{az},t}=\eta_{\mathrm{az},0}$.

Different from the BS-driven strategy, the angle tracking process in the UE-driven method is triggered at the UE side. Here, the direct ratio metric feedback is still applied, but the feedback process is configured by the UE according to the received signal strength corresponding to the anchor beam in the DDC. We explain the design procedures of the UE-driven angle tracking approach with the direct ratio metric feedback as follows:
\begin{itemize}
  \item For a given time-slot $t\in\left\{0,\cdots,T-1\right\}$ in the DTC, the auxiliary beam pair with the boresight angle identical to the steering direction of the anchor beam in the DDC is probed by the BS.
  \item The ratio metric $\zeta_{\mathrm{az},t}$ corresponding to the probed azimuth auxiliary beam pair is computed by the UE. Further, the received signal strength $\gamma_{t}$ of the anchor beam in the DDC is calculated by the UE. By comparing with the received signal strength $\gamma_{0}$ obtained at time-slot $0$, the received signal strength difference $\Delta\gamma_{t}=\left|\gamma_{t}-\gamma_{0}\right|$ is derived by the UE.
  \item The received signal strength difference $\Delta\gamma_{t}$ is compared with a predefined threshold $\varrho_{\mathrm{az}}$ such that if $\Delta\gamma_{t}\geq\varrho_{\mathrm{az}}$, the ratio metric $\zeta_{\mathrm{az},t}$ is quantized by the UE and sent back to the BS to trigger the anchor beam adjustment in the azimuth domain. Otherwise, the above process proceeds to time-slot $t+1$.
  \item Upon receiving $\zeta_{\mathrm{az},t}$, the BS estimates the channel's azimuth transmit spatial frequency $\hat{\psi}_{r^{\star},t}$ for time-slot $t$. The azimuth steering direction of the anchor beam in the DDC is then updated by the BS as $\eta_{\mathrm{az},t}=\hat{\psi}_{r^{\star},t}$.
\end{itemize}
Note that in the proposed UE-driven angle tracking with the direct ratio metric feedback, no prior knowledge of the auxiliary beam pair setup is required at the UE side, while only the received signal strength of the anchor beam is deduced as the triggering performance metric.

In the following, we illustrate the proposed angle tracking method with differential feedback from the perspective of the UE-driven design. The corresponding BS-driven differential feedback strategy can be similarly derived with moderate modifications on the tracking procedures.

\emph{2. UE-driven angle tracking design with differential ratio metric feedback}. To reduce the feedback overhead, we propose a differential ratio metric feedback option in this part. According to the derivation in (\ref{rmetric}), the ratio metric is distributed within $[-1,1]$. Further, the sign of the ratio metric implies the relative position of the angle to be estimated with respect to the boresight of the corresponding auxiliary beam pair. Consider the conceptual example shown in Fig.~4. For time-slot $t$, $\mathrm{sign}(\zeta_{\mathrm{az},t})=1$ indicates that $\psi_{r^{\star},t}$ falls on the left of the boresight of the corresponding auxiliary beam pair in the azimuth domain such that $\psi_{r^{\star},t}\in(\eta_{\mathrm{az},0}-\delta_{\mathrm{y}},\eta_{\mathrm{az},0})$. Similarly, $\psi_{r^{\star},t}\in(\eta_{\mathrm{az},0},\eta_{\mathrm{az},0}+\delta_{\mathrm{y}})$ results in $\mathrm{sign}(\zeta_{\mathrm{az},t})=-1$ implying that $\psi_{r^{\star},t}$ falls on the right of the boresight of the corresponding azimuth auxiliary beam pair.

At time-slot $t$, the UE derives the ratio metric $\zeta_{\mathrm{az},t}$ and calculates $\Delta\zeta_{\mathrm{az},t}=\left|\zeta_{\mathrm{az},t}-\zeta_{\mathrm{az},0}\right|$ and $\mathrm{sign}(\zeta_{\mathrm{az},t})$ for the azimuth domain. With the knowledge of the auxiliary beam pair setup, i.e., the boresight angle $\eta_{\mathrm{az},0}=\psi_{r^{\star},0}$, the boresight angle difference $\delta_{\mathrm{y}}$, and the corresponding beamwidth, the angle difference $\Delta\psi_{r^{\star},t}$ can be computed by the UE by exploiting the monotonic and symmetric properties of the ratio metric \cite{dztrans2d}. The angle difference is then compared with a predefined threshold $\varsigma_{\mathrm{az}}$ for the azimuth domain such that if $\Delta\psi_{r^{\star},t}\geq\varsigma_{\mathrm{az}}$, the UE quantizes $\Delta\zeta_{\mathrm{az},t}$ and sends it back to the BS along with $\mathrm{sign}(\zeta_{\mathrm{az},t})$ to trigger the anchor beam adjustment. Note that in contrast to the direct ratio metric quantization, the differential ratio metric quantization reduces the feedback overhead by half but with one extra bit indicating the sign. Upon receiving the feedback information, the BS can determine $\Delta\psi_{r^{\star},t}$ using $\Delta\zeta_{\mathrm{az},t}$. The azimuth steering direction of the anchor beam in the DDC can therefore be updated as $\eta_{\mathrm{az},t}=\hat{\psi}_{r^{\star},t}=\psi_{r^{\star},0}+\mathrm{sign}(\zeta_{\mathrm{az},t})\Delta\psi_{r^{\star},t}$.

\textbf{Remark}: Similar to the direct and differential ratio metric feedback methods, direct and differential angle feedback strategies can also be supported for the angle tracking designs, as long as necessary auxiliary beam pair setup is available at the UE side.

\section{Impact of Radiation Pattern Impairments}
Because of manufacturing inaccuracies, a variety of impairments such as geometrical and electrical tolerances cause non-uniform amplitude and phase characteristics of the individual antenna elements \cite{gsdhrk}. This results in phase and amplitude errors of the radiation patterns \cite{bcng}. In this paper, we first define the following three terminologies:
\begin{itemize}
  \item \emph{Ideal radiation pattern}: the radiation pattern is not impaired by any types of impairments such as phase and amplitude errors, mutual coupling, imperfect matching, and etc.
  \item \emph{Impaired radiation pattern}: the radiation pattern is impaired only by the phase and amplitude errors, but not other impairments such as mutual coupling, imperfect matching, and etc.
  \item \emph{Calibrated radiation pattern}: the phase and amplitude errors are compensated by certain array calibration methods; after the calibration, residual phase and amplitude errors may still exist depending on many factors, e.g., the calibration SNR or distribution of the impairments.
\end{itemize}
The angle tracking performance of the proposed auxiliary beam pair-assisted designs is subject to the radiation pattern impairments, which are neglected during the derivation of the ratio metric in (\ref{rmetric}). If the radiation patterns of the beams in the auxiliary beam pair are impaired by the phase and amplitude errors, the monotonic and symmetric properties of the ratio metric may not hold, which in turn, results in large angle tracking errors. In the following, we first illustrate the impact of the radiation pattern impairments on the proposed angle tracking designs. To calibrate the antenna array with the analog architecture, we custom design and evaluate two calibration methods. We then examine the impact of the residual calibration errors on the proposed angle tracking approaches.

\subsection{Impact of phase and amplitude errors on proposed methods}
Neglecting mutual coupling and matching effects, and denoting the phase and amplitude error matrices by $\bm{P}$ and $\bm{A}$, we have $\bm{P}=\mathrm{diag}\left(\left[e^{\mathrm{j}p_0},e^{\mathrm{j}p_1},\cdots,e^{\mathrm{j}p_{N_{\mathrm{tot}}-1}}\right]^{\mathrm{T}}\right)$ and $\bm{A}=\mathrm{diag}\Big(\big[a_0,a_1,\cdots,\\a_{N_{\mathrm{tot}}-1}\big]^{\mathrm{T}}\Big)$, where $p_{i}$ and $a_{i}$ correspond to the phase and amplitude errors on the $i$-th antenna element with $i=0,\cdots,N_{\mathrm{tot}}-1$. Due to the UPA structure, we can decompose $\bm{P}$ and $\bm{A}$ as $\bm{P}=\bm{P}_{\mathrm{el}}\otimes\bm{P}_{\mathrm{az}}$ and $\bm{A}=\bm{A}_{\mathrm{el}}\otimes\bm{A}_{\mathrm{az}}$, where $\bm{P}_{\mathrm{el}}=\mathrm{diag}\left(\left[e^{\mathrm{j}p_{\mathrm{el},0}},e^{\mathrm{j}p_{\mathrm{el},1}},\cdots,e^{\mathrm{j}p_{\mathrm{el},N_{\mathrm{x}}-1}}\right]^{\mathrm{T}}\right)$ and $\bm{A}_{\mathrm{el}}=\mathrm{diag}\left(\left[a_{\mathrm{el},0},a_{\mathrm{el},1},\cdots,a_{\mathrm{el},N_{\mathrm{x}}-1}\right]^{\mathrm{T}}\right)$ correspond to the elevation domain, and $\bm{P}_{\mathrm{az}}=\mathrm{diag}\Big(\big[e^{\mathrm{j}p_{\mathrm{az},0}},\\e^{\mathrm{j}p_{\mathrm{az},1}},\cdots,e^{\mathrm{j}p_{\mathrm{az},N_{\mathrm{y}}-1}}\big]^{\mathrm{T}}\Big)$ and $\bm{A}_{\mathrm{az}}=\mathrm{diag}\left(\left[a_{\mathrm{az},0},a_{\mathrm{az},1},\cdots,a_{\mathrm{az},N_{\mathrm{y}}-1}\right]^{\mathrm{T}}\right)$ are for the azimuth domain. In this paper, we model $p_{\mathrm{el},i_{\mathrm{el}}}$, $a_{\mathrm{el},i_{\mathrm{el}}}$ with $i_{\mathrm{el}}=0,\cdots,N_{\mathrm{x}}-1$ and $p_{\mathrm{az},i_{\mathrm{az}}}$, $a_{\mathrm{az},i_{\mathrm{az}}}$ with $i_{\mathrm{az}}=0,\cdots,N_{\mathrm{y}}-1$ as Gaussian distributed random variables with zero mean and certain variances.

We employ the example shown in Fig.~3(a) to illustrate the impact of the phase and amplitude errors on the auxiliary beam pair design. Denote $\bm{C}=\bm{A}\bm{P}$ and neglect the radiation pattern impairments at the UE side. Using the transmit analog beam $\bm{a}_{\mathrm{t}}(\eta_{\mathrm{el}},\eta_{\mathrm{az}}-\delta_{\mathrm{y}})$ and the receive analog beam $\bm{a}_{\mathrm{r}}(\vartheta)$, we compute the corresponding noiseless received signal strength as
\begin{eqnarray}
\chi^{\Delta}_{\mathrm{az}}&=&\left|g_{r^{\star}}\bar{\rho}_{\tau_{r^{\star}}}\right|^{2}\left|\bm{a}^{*}_{\mathrm{r}}(\vartheta)\bm{a}_{\mathrm{r}}(\nu_{r^{\star}})\right|^{2}\nonumber\\
&\times&\bm{a}^{*}_{\mathrm{t}}(\eta_{\mathrm{el}},\eta_{\mathrm{az}}-\delta_{\mathrm{y}})\bm{C}^{*}\bm{a}_{\mathrm{t}}(\theta_{r^{\star}},\psi_{r^{\star}})\bm{a}_{\mathrm{t}}^{*}(\theta_{r^{\star}},\psi_{r^{\star}})\bm{C}\bm{a}_{\mathrm{t}}(\eta_{\mathrm{el}},\eta_{\mathrm{az}}-\delta_{\mathrm{y}})\\
&=&\left|g_{r^{\star}}\bar{\rho}_{\tau_{r^{\star}}}\right|^{2}\left|\bm{a}^{*}_{\mathrm{r}}(\vartheta)\bm{a}_{\mathrm{r}}(\nu_{r^{\star}})\right|^{2}\nonumber\\
&\times&\left|\sum_{i_{\mathrm{el}}=0}^{N_{\mathrm{x}}-1}a_{\mathrm{el},i_{\mathrm{el}}}e^{-\mathrm{j}\left[i_{\mathrm{el}}(\theta_{r^{\star}}-\eta_{\mathrm{el}})-p_{\mathrm{el},i_{\mathrm{el}}}\right]}\right|^2\left|\sum_{i_{\mathrm{az}}=0}^{N_{\mathrm{y}}-1}a_{\mathrm{az},i_{\mathrm{az}}}e^{-\mathrm{j}\left[i_{\mathrm{az}}(\psi_{r^{\star}}-\eta_{\mathrm{az}}+\delta_{\mathrm{y}})-p_{\mathrm{az},i_{\mathrm{az}}}\right]}\right|^2.\label{newcal}
\end{eqnarray}
Similarly, we can derive the received signal strength with respect to the transmit and receive beams pair $\bm{a}_{\mathrm{t}}(\eta_{\mathrm{el}},\eta_{\mathrm{az}}+\delta_{\mathrm{y}})$ and $\bm{a}_{\mathrm{r}}(\vartheta)$ as
\begin{eqnarray}
\chi^{\Sigma}_{\mathrm{az}}&=&\left|g_{r^{\star}}\bar{\rho}_{\tau_{r^{\star}}}\right|^{2}\left|\bm{a}^{*}_{\mathrm{r}}(\vartheta)\bm{a}_{\mathrm{r}}(\nu_{r^{\star}})\right|^{2}\nonumber\\
&\times&\left|\sum_{i_{\mathrm{el}}=0}^{N_{\mathrm{x}}-1}a_{\mathrm{el},i_{\mathrm{el}}}e^{-\mathrm{j}\left[i_{\mathrm{el}}(\theta_{r^{\star}}-\eta_{\mathrm{el}})-p_{\mathrm{el},i_{\mathrm{el}}}\right]}\right|^2\left|\sum_{i_{\mathrm{az}}=0}^{N_{\mathrm{y}}-1}a_{\mathrm{az},i_{\mathrm{az}}}e^{-\mathrm{j}\left[i_{\mathrm{az}}(\psi_{r^{\star}}-\eta_{\mathrm{az}}-\delta_{\mathrm{y}})-p_{\mathrm{az},i_{\mathrm{az}}}\right]}\right|^2.\label{newcal1}
\end{eqnarray}
Due to the phase and amplitude errors, the received signal strengths $\chi^{\Delta}_{\mathrm{az}}$ and $\chi^{\Sigma}_{\mathrm{az}}$ in (\ref{newcal}) and (\ref{newcal1}) can not be expressed as the same forms as those in (21) and (\ref{expaaaa}). The corresponding ratio metric calculated via $\zeta_{\mathrm{az}}=\frac{\chi^{\Delta}_{\mathrm{az}}-\chi^{\Sigma}_{\mathrm{az}}}{\chi^{\Delta}_{\mathrm{az}}+\chi^{\Sigma}_{\mathrm{az}}}$ is therefore no longer a strict monotonic function of the angle to be estimated. By directly inverting the ratio metric function as according to (\ref{angleestimate}), high angle estimation error probability could be incurred, which in turn, degrades the angle tracking performance of the proposed methods.

\begin{figure}
\centering
\subfigure[]{%
\includegraphics[width=2.63in]{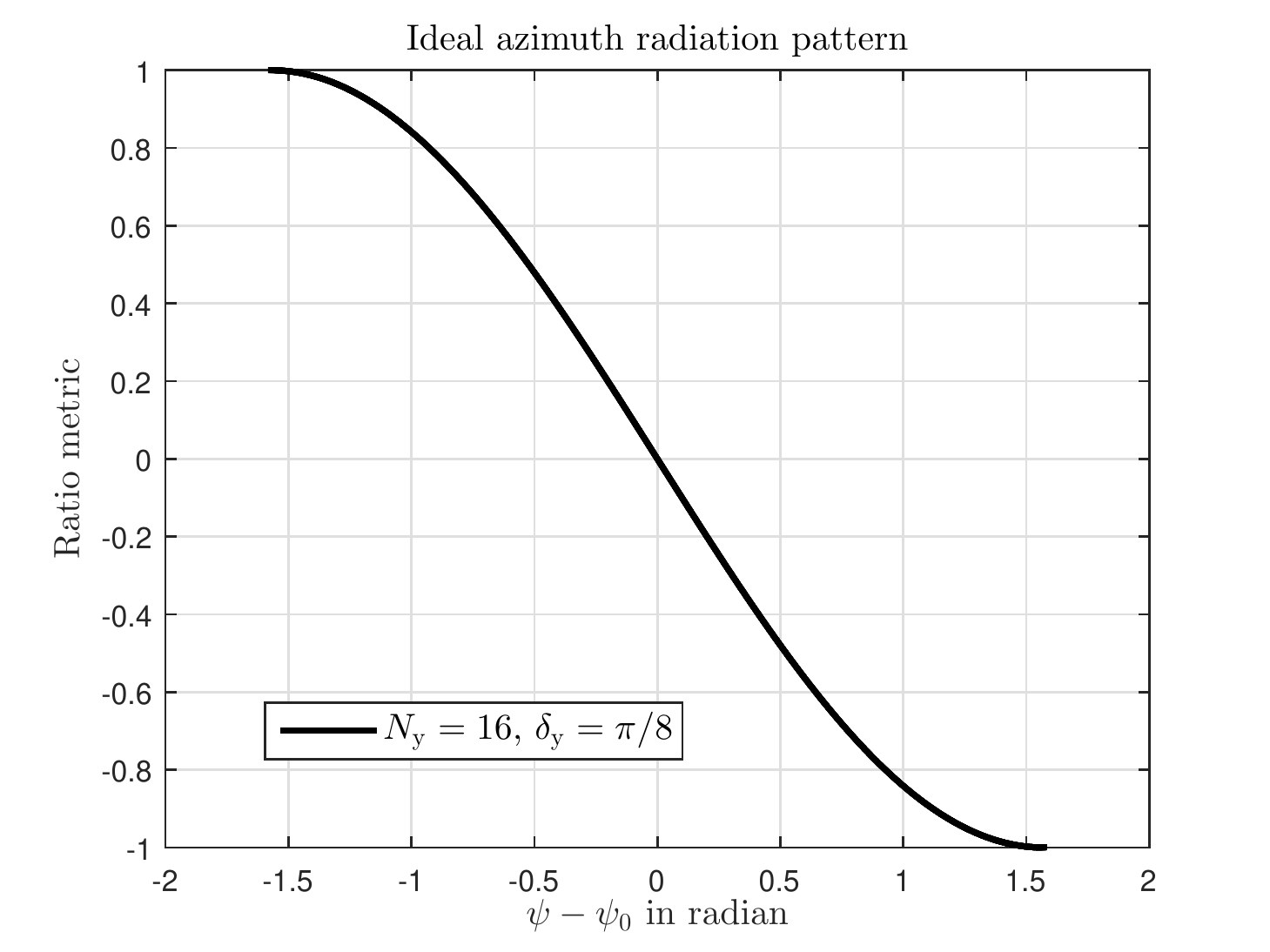}
\label{fig:subfigure1}}
\quad
\subfigure[]{%
\includegraphics[width=2.63in]{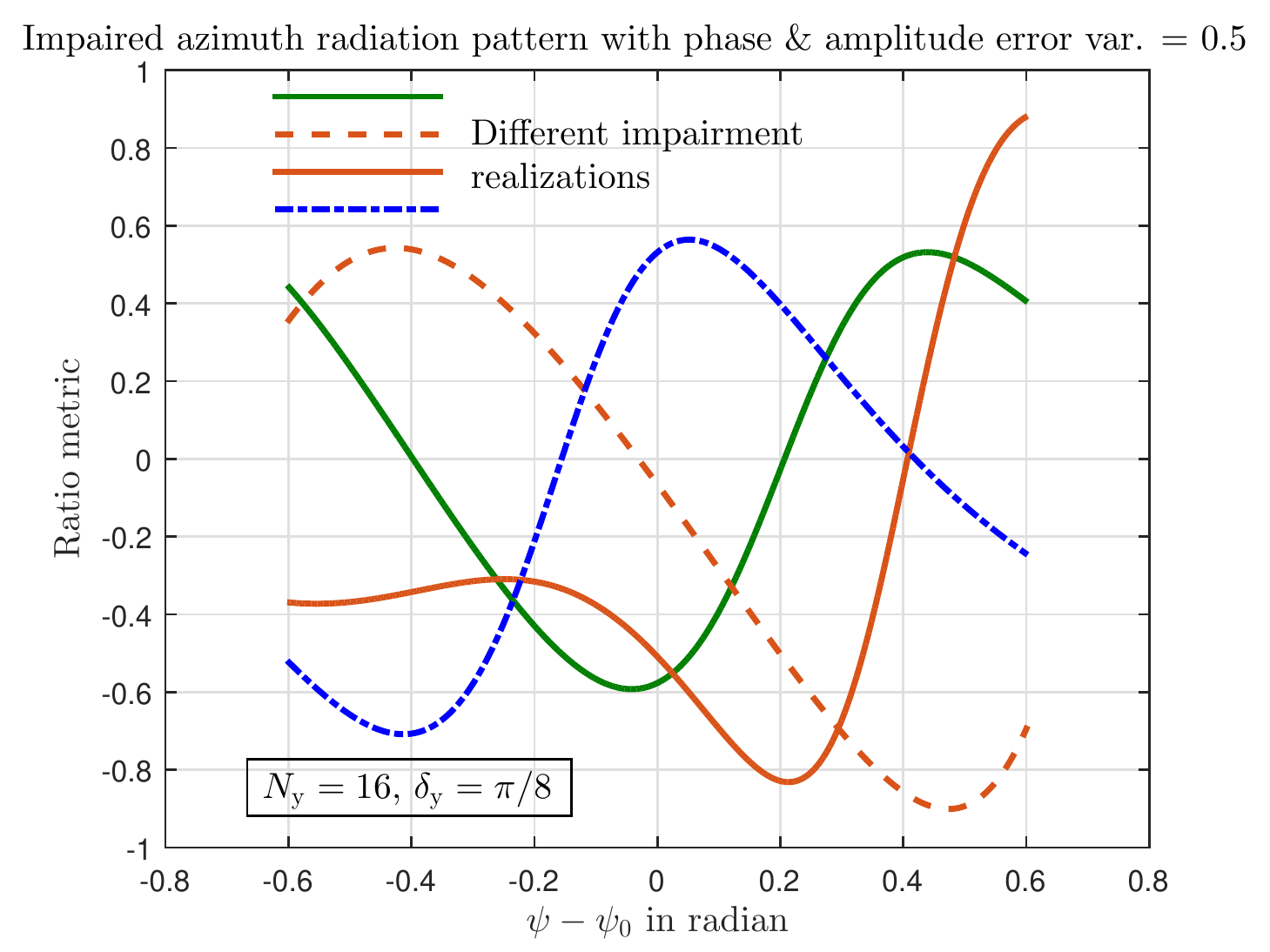}
\label{fig:subfigure2}}
\caption{(a) Ratio metric versus angles to be estimated under ideal azimuth radiation pattern with $N_{\mathrm{y}}=16$ and $\delta_{\mathrm{y}}=\pi/8$. (b) Ratio metrics versus angles to be estimated under different realizations of impaired azimuth radiation patterns with $N_{\mathrm{y}}=16$ and $\delta_{\mathrm{y}}=\pi/8$. The phase $\&$ amplitude errors variances are $0.5$.}
\label{fig:figure}
\end{figure}
In Figs.~5(a) and (b), we plot the ratio metrics versus the angle to be estimated assuming ideal radiation pattern and impaired radiation pattern with $N_{\mathrm{y}}=16$ and $\delta_{\mathrm{y}}=\pi/8$. It is observed from Fig.~5(b) that with $0.5$ phase and amplitude errors variances, the ratio metrics obtained via different impairment realizations are neither monotonic functions of the angle to be estimated nor symmetrical with respect to the origin. These observations are consistent with our analysis. Practical implementation of the proposed angle tracking designs therefore requires array calibration to compensate for the phase and amplitude errors.

Conventional array calibration methods such as those in \cite{arrayca} can not be directly applied to the array setup shown in Fig.~1. This is because in the employed array architecture, all antenna elements are driven by a limited number of RF chains such that only $N_{\mathrm{RF}}$-dimensional measurements are accessible to calibrate all $N_{\mathrm{tot}}$ antenna elements. In this paper, we develop and evaluate two off-line array calibration methods for the employed array configurations assuming simple LOS channels and single-carrier setup.

\subsection{Receive combining based array calibration with single calibration source}
In this method, we assume that the single calibration source transmitting the calibration reference signal (RS) is located at the origin with respect to the BS antenna array such that the calibration RS impinges on the antenna array at $0$ degree in both the azimuth and elevation domains. At the BS, a set of receive combining vectors are formed in a time-division multiplexing (TDM) manner probing towards $N_{\mathrm{tot}}$ different angular directions in both the azimuth and elevation domains. The external calibration source can be placed close to the BS antenna array, and the channel between them is LOS. We can therefore express the signals received across all the $N_{\mathrm{tot}}$ receive probings as
\begin{eqnarray}
y_0 &=& \bm{a}_{\mathrm{t}}^{*}(\eta_{\mathrm{el},0},\eta_{\mathrm{az},0})\bm{C}\bm{a}_{\mathrm{t}}(\theta,\psi)x+\bm{a}_{\mathrm{t}}^{*}(\eta_{\mathrm{el},0},\eta_{\mathrm{az},0})\bm{n}_0\\
&\vdots&\nonumber\\
y_{N_{\mathrm{tot}}-1} &=& \bm{a}_{\mathrm{t}}^{*}(\eta_{\mathrm{el},N_{\mathrm{x}}-1},\eta_{\mathrm{az},N_{\mathrm{y}}-1})\bm{C}\bm{a}_{\mathrm{t}}(\theta,\psi)x+\bm{a}_{\mathrm{t}}^{*}(\eta_{\mathrm{el},N_{\mathrm{x}}-1},\eta_{\mathrm{az},N_{\mathrm{y}}-1})\bm{n}_{N_{\mathrm{tot}}-1},
\end{eqnarray}
where $x$ represents the calibration RS, $\theta=\psi=0$, $\eta_{\mathrm{el},i_{\mathrm{el}}}$ and $\eta_{\mathrm{az},i_{\mathrm{az}}}$ ($i_{\mathrm{el}}=0,\cdots,N_{\mathrm{x}}-1$ and $i_{\mathrm{az}}=0,\cdots,N_{\mathrm{y}}-1$) are the receive steering directions in the elevation and azimuth domains, and $\bm{n}_{i}$ ($i=0,\cdots,N_{\mathrm{tot}}-1$) is the corresponding noise vector. In this paper, we assume the calibration RS $x=1$ while it can be selected as a different symbol from $1$ as long as it is known a prior. By concatenating all the received signal samples $y_0,\cdots,y_{N_{\mathrm{tot}}-1}$, we therefore have
\begin{equation}\label{cali}
\bm{y}=\left[
         \begin{array}{c}
           y_0 \\
           \vdots \\
           y_{N_{\mathrm{tot}}-1} \\
         \end{array}
       \right]=\bm{A}_{\mathrm{t}}\bm{C}\bm{1}_{N_{\mathrm{tot}}\times1}+\bm{A}_{\mathrm{t}}\bm{n},
\end{equation}
with
\begin{eqnarray}\label{a0}
\bm{A}_{\mathrm{t}}=\left[
         \begin{array}{c}
           \bm{a}_{\mathrm{t}}^{*}(\eta_{\mathrm{el},0},\eta_{\mathrm{az},0}) \\
           \vdots \\
           \bm{a}_{\mathrm{t}}^{*}(\eta_{\mathrm{el},N_{\mathrm{x}}-1},\eta_{\mathrm{az},N_{\mathrm{y}}-1}) \\
         \end{array}
       \right]\in\mathbb{C}^{N_{\mathrm{tot}}\times N_{\mathrm{tot}}},\hspace{3mm}\bm{n}=\left[
         \begin{array}{c}
           \bm{n}_0 \\
           \vdots \\
           \bm{n}_{N_{\mathrm{tot}}-1} \\
         \end{array}
       \right]\in\mathbb{C}^{N_{\mathrm{tot}}\times 1}.
\end{eqnarray}
According to (\ref{cali}), the phase and amplitude errors matrix can be estimated as
\begin{equation}
\hat{\bm{C}}=\mathrm{diag}\left\{\bm{A}_{\mathrm{t}}^{-1}\bm{y}\right\},
\end{equation}
and the calibration matrix is determined as $\bm{K}=\hat{\bm{C}}^{-1}$. Note that with different receive steering directions and DFT-type receive steering vector structure, the square matrix $\bm{A}_{\mathrm{t}}$ is invertible.

\subsection{Receive combining based array calibration with distributed calibration sources}
In this method, a total of $N_{\mathrm{RS}}$ distributed calibration sources transmit incoherent calibration RSs to the BS antenna array. Different from the single calibration source case, a total of $N_{\mathrm{RF}}$ receive beams are simultaneously probed by the BS to receive the calibration RSs in the TDM round-robing fashion. Calibrating all the antenna elements therefore requires $N_{\mathrm{RS}}=N_{\mathrm{tot}}/N_{\mathrm{RF}}$. We can then express the received signal model as
\begin{equation}
\bm{Y}=\underline{\bm{A}}_{\mathrm{t}}\bm{C}\bm{B}_{\mathrm{t}}\bm{I}_{N_{\mathrm{RS}}}x + \underline{\bm{A}}_{\mathrm{t}}\bm{N},
\end{equation}
where $\bm{Y}\in\mathbb{C}^{N_{\mathrm{RF}}\times N_{\mathrm{RS}}}$,
\begin{eqnarray}\label{a1}
\underline{\bm{A}}_{\mathrm{t}}=\left[
         \begin{array}{c}
           \bm{a}_{\mathrm{t}}^{*}(\eta_{\mathrm{el},0},\eta_{\mathrm{az},0}) \\
           \vdots \\
           \bm{a}_{\mathrm{t}}^{*}(\eta_{\mathrm{el},N_{\mathrm{RF}}-1},\eta_{\mathrm{az},N_{\mathrm{RF}}-1}) \\
         \end{array}
       \right],\hspace{3mm}\bm{B}_{\mathrm{t}}=\left[\bm{a}_{\mathrm{t}}(\theta_0,\psi_0),\hspace{2mm}\cdots,\hspace{2mm}\bm{a}_{\mathrm{t}}(\theta_{N_{\mathrm{RS}}-1},\psi_{N_{\mathrm{RS}}-1})\right],
\end{eqnarray}
and $\bm{N}$ represents the $N_{\mathrm{tot}}\times N_{\mathrm{RS}}$ noise matrix. Note that because the calibration is conducted off-line, the receive steering matrix $\underline{\bm{A}}_{\mathrm{t}}$ and the array response matrix $\bm{B}_{\mathrm{t}}$ are known a prior, which can be used to determine $\bm{V}_{\mathrm{t}}=\bm{B}^{\mathrm{T}}_{\mathrm{t}}\otimes\underline{\bm{A}}_{\mathrm{t}}$. Assuming $x=1$, the phase and amplitude errors matrix can then be estimated as
\begin{equation}
\hat{\bm{C}}=\mathrm{diag}\left(\left(\bm{V}^{*}_{\mathrm{t}}\bm{V}_{\mathrm{t}}\right)^{-1}\bm{V}^{*}_{\mathrm{t}}\mathrm{vec}(\bm{Y})\right).
\end{equation}
We can therefore calculate the calibration matrix $\bm{K}$ as $\bm{K}=\hat{\bm{C}}^{-1}$.

\begin{figure}
\centering
\subfigure[]{%
\includegraphics[width=2.63in]{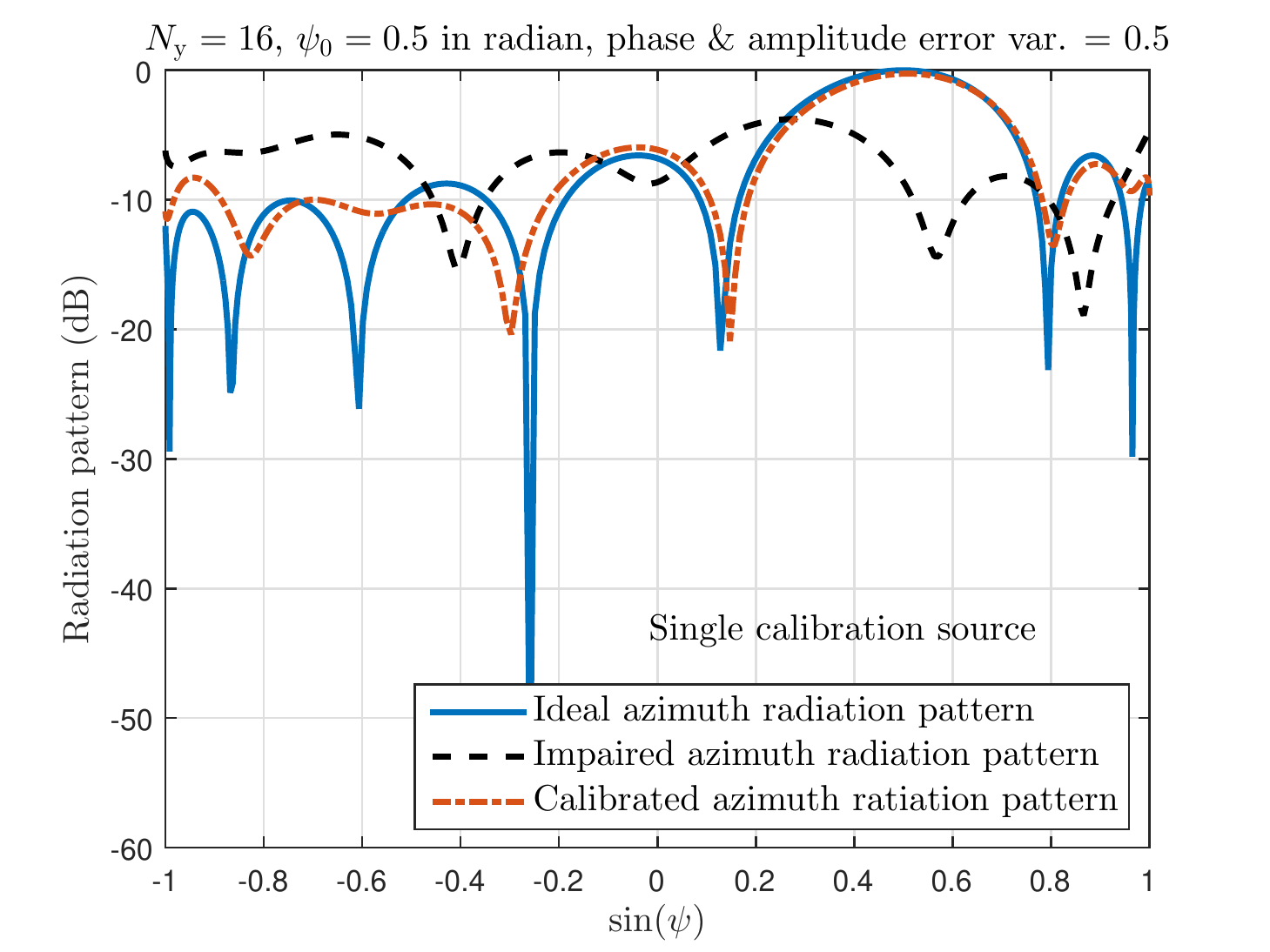}
\label{fig:subfigure1}}
\quad
\subfigure[]{%
\includegraphics[width=2.63in]{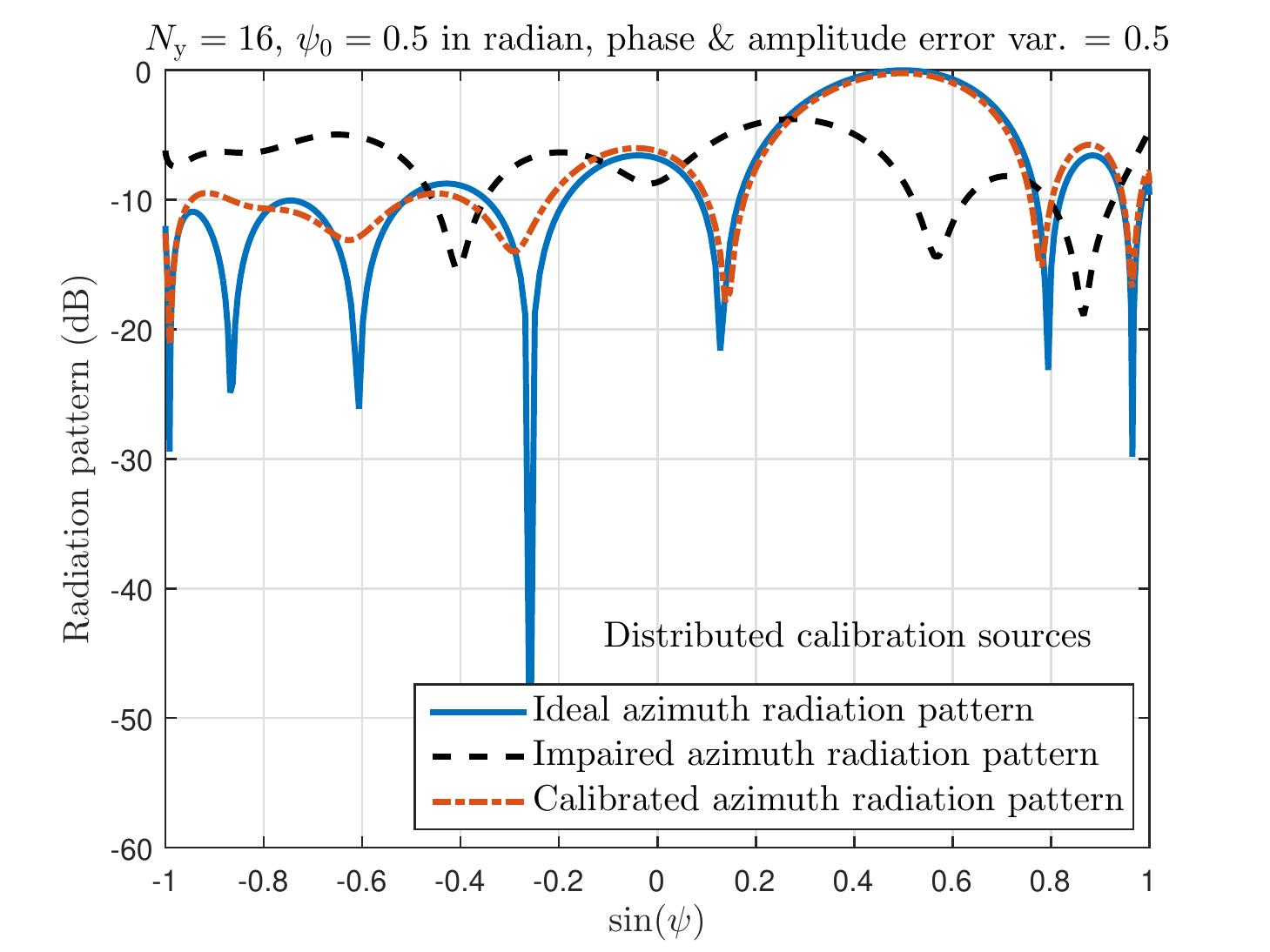}
\label{fig:subfigure2}}
\subfigure[]{%
\includegraphics[width=2.63in]{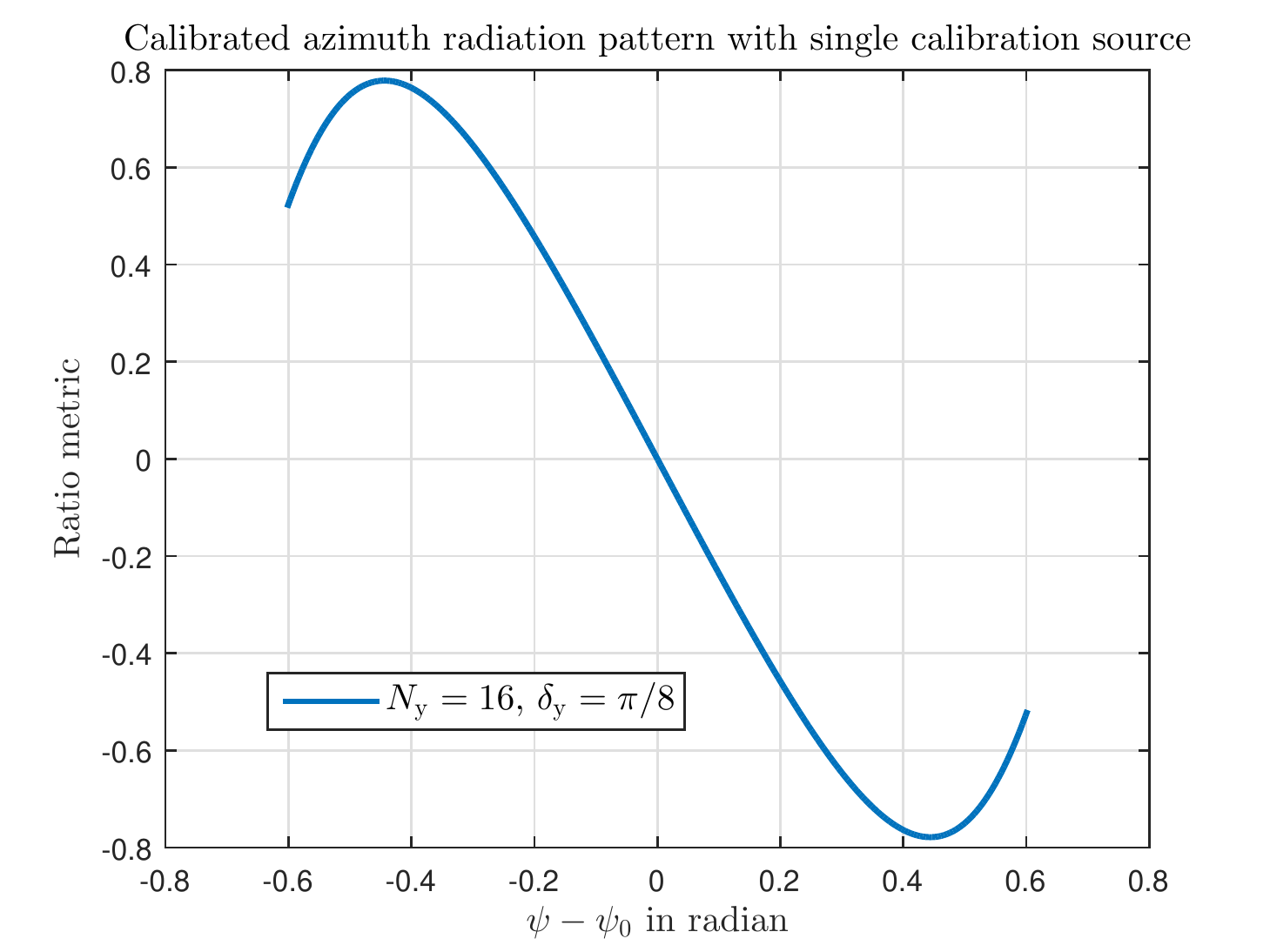}
\label{fig:subfigure1}}
\quad
\subfigure[]{%
\includegraphics[width=2.63in]{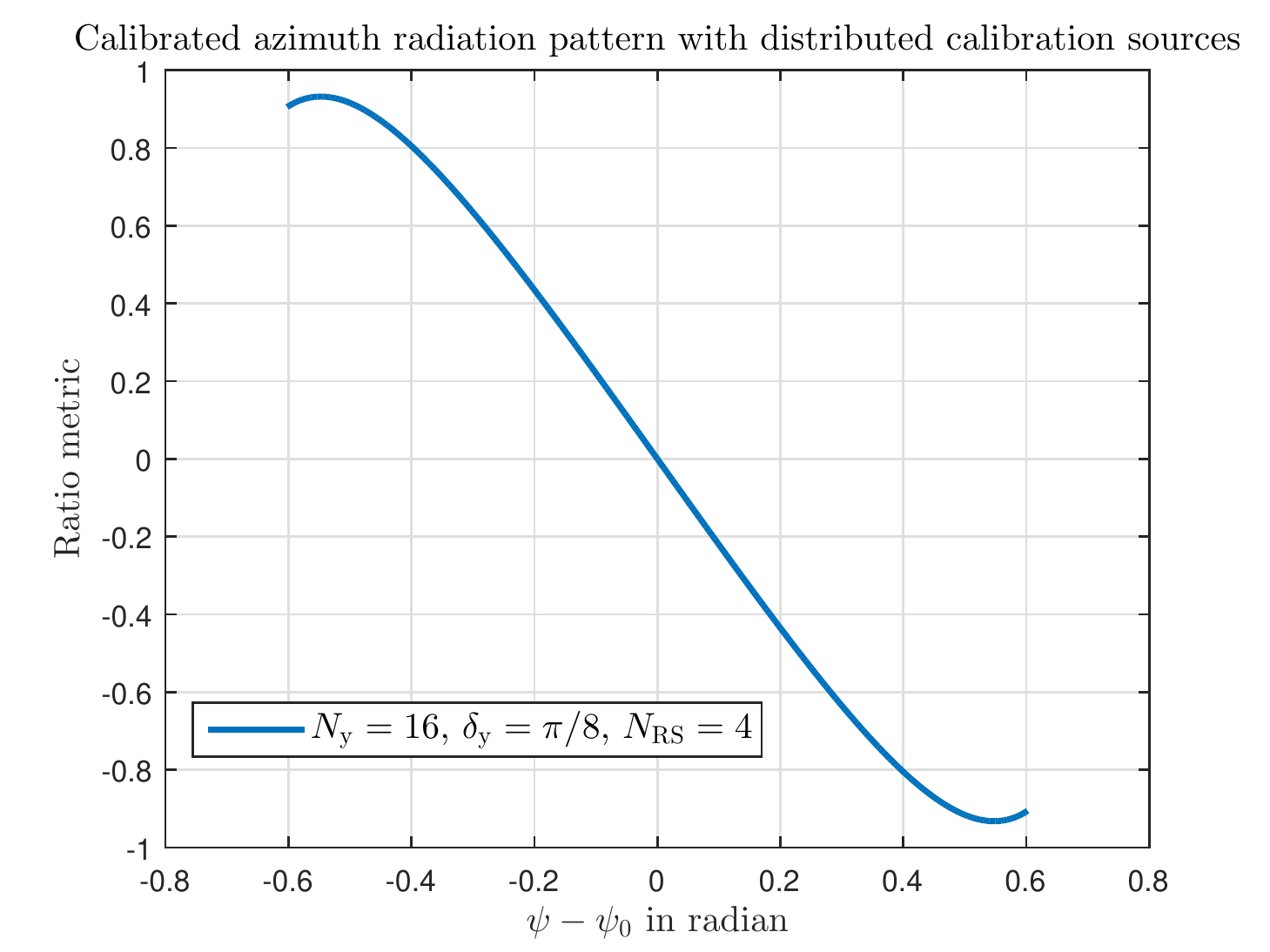}
\label{fig:subfigure2}}
\caption{(a) Calibrated azimuth radiation pattern using the proposed single calibration source and receive combining based calibration method; $N_{\mathrm{x}}=1$, $N_{\mathrm{y}}=16$ with the phase $\&$ amplitude errors variances $0.5$. (b) Calibrated azimuth radiation pattern using the proposed distributed calibration sources and receive combining based calibration method; $N_{\mathrm{x}}=1$, $N_{\mathrm{y}}=16$ with the phase $\&$ amplitude errors variances $0.5$; $N_{\mathrm{RF}}=4$ and $N_{\mathrm{RS}}=4$. (c) Ratio metric versus angles to be estimated under calibrated azimuth radiation pattern with the proposed single calibration source based method; $N_{\mathrm{x}}=1$, $N_{\mathrm{y}}=16$ and $\delta_{\mathrm{y}}=\pi/8$; the phase $\&$ amplitude errors variances are $0.5$. (d) Ratio metric versus angles to be estimated under calibrated azimuth radiation pattern with the proposed distributed calibration sources based method; $N_{\mathrm{x}}=1$, $N_{\mathrm{y}}=16$ and $\delta_{\mathrm{y}}=\pi/8$; the phase $\&$ amplitude errors variances are $0.5$; $N_{\mathrm{RF}}=4$ and $N_{\mathrm{RS}}=4$.}
\label{fig:figure}
\end{figure}
In Figs.~6(a) and (b), we evaluate the impact of the residual calibration errors on the azimuth radiation patterns for the proposed calibration methods. We set the calibration SNR as $0$ dB. As can be seen from Figs.~6(a) and (b), the calibrated radiation patterns almost match with the ideal radiation patterns in the azimuth domain such that the main lobe and side lobes can be clearly differentiated. Note that with increase in the calibration SNR, the calibration performances can be further improved.

After the array calibration, the amplitude and phase errors become small and are approximately the same across all the antenna elements. Denote the residual amplitude and phase errors for all antennas by $\bar{a}$ and $\bar{p}$. Using the transmit analog beam $\bm{a}_{\mathrm{t}}(\eta_{\mathrm{el}},\eta_{\mathrm{az}}-\delta_{\mathrm{y}})$ and the receive analog beam $\bm{a}_{\mathrm{r}}(\vartheta)$, we can obtain the noiseless received signal strength as
\begin{eqnarray}
\chi^{\Delta}_{\mathrm{az}}&=&\left|g_{r^{\star}}\bar{\rho}_{\tau_{r^{\star}}}\right|^{2}\left|\bm{a}^{*}_{\mathrm{r}}(\vartheta)\bm{a}_{\mathrm{r}}(\nu_{r^{\star}})\right|^{2}\left|\bm{a}_{\mathrm{t}}^{*}(\theta_{r^{\star}},\psi_{r^{\star}})\bm{C}\bm{K}\bm{a}_{\mathrm{t}}(\eta_{\mathrm{el}},\eta_{\mathrm{az}}-\delta_{\mathrm{y}})\right|^2\\
&\approx&\left|g_{r^{\star}}\bar{\rho}_{\tau_{r^{\star}}}\right|^{2}\left|\bm{a}^{*}_{\mathrm{r}}(\vartheta)\bm{a}_{\mathrm{r}}(\nu_{r^{\star}})\right|^{2}\left|\bar{a}e^{\mathrm{j}\bar{p}}\right|^2\left|\sum_{i_{\mathrm{el}}=0}^{N_{\mathrm{x}}-1}e^{-\mathrm{j}i_{\mathrm{el}}(\theta_{r^{\star}}-\eta_{\mathrm{el}})}\right|^2\left|\sum_{i_{\mathrm{az}}=0}^{N_{\mathrm{y}}-1}e^{-\mathrm{j}i_{\mathrm{az}}(\psi_{r^{\star}}-\eta_{\mathrm{az}}+\delta_{\mathrm{y}})}\right|^2\nonumber\\
\\
&=&\left|g_{r^{\star}}\bar{\rho}_{\tau_{r^{\star}}}\right|^{2}\left|\bm{a}^{*}_{\mathrm{r}}(\vartheta)\bm{a}_{\mathrm{r}}(\nu_{r^{\star}})\right|^{2}\left|\bar{a}e^{\mathrm{j}\bar{p}}\right|^2\frac{\sin^{2}\left(\frac{N_{\mathrm{x}}(\theta_{r^{\star}}-\eta_{\mathrm{el}})}{2}\right)}{\sin^{2}\left(\frac{\theta_{r^{\star}}-\eta_{\mathrm{el}}}{2}\right)}\frac{\sin^{2}\left(\frac{N_{\mathrm{y}}(\psi_{r^{\star}}-\eta_{\mathrm{az}})}{2}\right)}{\sin^{2}\left(\frac{\psi_{r^{\star}}-\eta_{\mathrm{az}}+\delta_{\mathrm{y}}}{2}\right)}.
\end{eqnarray}
Similarly, after the array calibration, we can compute the received signal strength with respect to the transmit and receive beams pair $\bm{a}_{\mathrm{t}}(\eta_{\mathrm{el}},\eta_{\mathrm{az}}+\delta_{\mathrm{y}})$ and $\bm{a}_{\mathrm{r}}(\vartheta)$ as
\begin{eqnarray}\label{newcal1sss}
\chi^{\Sigma}_{\mathrm{az}}&\approx&\left|g_{r^{\star}}\bar{\rho}_{\tau_{r^{\star}}}\right|^{2}\left|\bm{a}^{*}_{\mathrm{r}}(\vartheta)\bm{a}_{\mathrm{r}}(\nu_{r^{\star}})\right|^{2}\left|\bar{a}e^{\mathrm{j}\bar{p}}\right|^2\frac{\sin^{2}\left(\frac{N_{\mathrm{x}}(\theta_{r^{\star}}-\eta_{\mathrm{el}})}{2}\right)}{\sin^{2}\left(\frac{\theta_{r^{\star}}-\eta_{\mathrm{el}}}{2}\right)}\frac{\sin^{2}\left(\frac{N_{\mathrm{y}}(\psi_{r^{\star}}-\eta_{\mathrm{az}})}{2}\right)}{\sin^{2}\left(\frac{\psi_{r^{\star}}-\eta_{\mathrm{az}}-\delta_{\mathrm{y}}}{2}\right)}.
\end{eqnarray}
The corresponding ratio metric derived by $\zeta_{\mathrm{az}}=\frac{\chi^{\Delta}_{\mathrm{az}}-\chi^{\Sigma}_{\mathrm{az}}}{\chi^{\Delta}_{\mathrm{az}}+\chi^{\Sigma}_{\mathrm{az}}}$ therefore exhibits the same form as (\ref{rmetric}) implying that the channel directional information can be retrieved by inverting the ratio metric. In Figs.~6(c) and (d), we plot the ratio metrics obtained after the array calibration with respect to the angle to be estimated in the azimuth domain. We evaluate both the single calibration source and distributed calibration sources based methods. By comparing Figs.~6(c) and (d) with Fig.~5(a), it can be observed that the monotonic and symmetric properties of the ratio metric hold for most of the angle values.
\section{Numerical Results}
\begin{figure}
\centering
\subfigure[]{%
\includegraphics[width=2.55in]{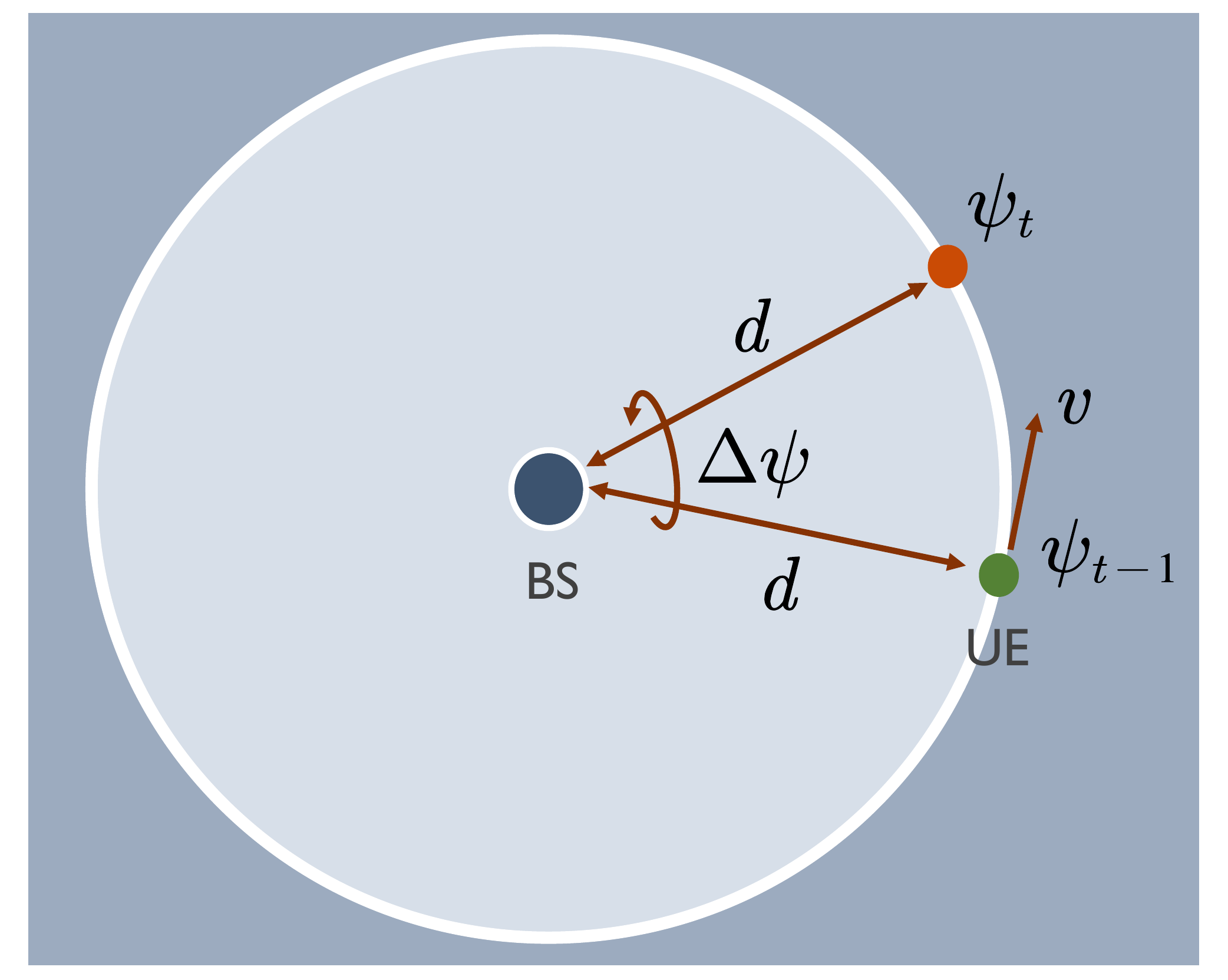}
\label{fig:subfigure1}}
\quad
\quad
\subfigure[]{%
\includegraphics[width=2.55in]{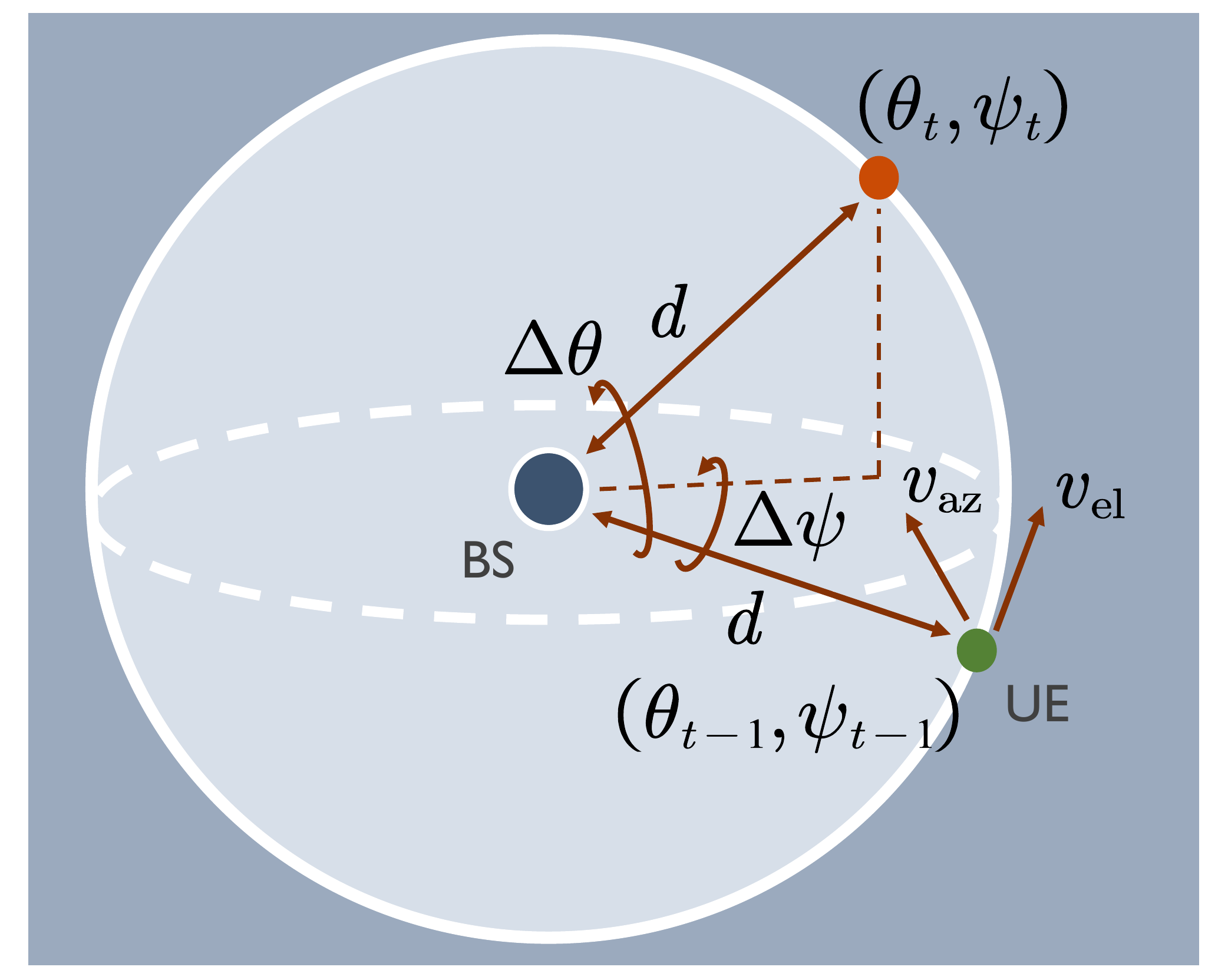}
\label{fig:subfigure2}}
\caption{(a) Angular motion model I. The BS is located at the origin of a ring, and the UE is moving along the ring with certain absolute speed $v$. The radius of the ring is denoted by $d$. (b) Angular motion model II. The BS is located at the origin of a sphere, and the UE is moving on the surface with certain absolute speeds $v_{\mathrm{az}}$ and $v_{\mathrm{el}}$ towards the azimuth and elevation domains. The radius of the sphere is denoted by $d$.}
\label{fig:figure}
\end{figure}
In this section, we evaluate the proposed BS-driven angle tracking design with the direct ratio metric feedback and the periodic DTC. Note that different angle tracking strategies developed in Section III exhibit similar tracking performances, though they have different requirements on the tracking triggering metric, feedback information, and information available at the UE side. We evaluate the proposed angle tracking method assuming ideal radiation pattern, impaired radiation pattern with phase and amplitude errors, and calibrated radiation pattern. For simplicity, we obtain the calibrated radiation pattern via the proposed single calibration source based strategy. We set the angle difference threshold for triggering the beam adjustment as $10^{\circ}$. As the ratio metric is non-uniformly distributed within the interval of $[-1,1]$ \cite{dztrans}, we employ the Lloyd's algorithm \cite{lollds} to optimize the codebook for quantizing the ratio metric.

\subsection{Narrowband single-path channels with single-carrier}
In this part, we provide the numerical results in narrowband single-path channels with single-carrier modulation. We consider a single UE and two angular motion models shown in Figs.~7(a) and (b) to reveal the moving trajectory of the UE. In the first model (angular motion model I), the ULA is employed at the BS, while in the second model (angular motion model II), the UPA is employed at the BS such that the tracking beams can be probed towards both the elevation and azimuth domains. For both cases, the ULA is assumed at the UE side. Note that we develop the angular motion models I and II to better characterize the angle variations in terms of the moving trajectory. In Section V-B, we employ statistical temporal evolution tools to model practical channel variations. We list other simulation assumptions and parameters in Table I. We drop the path index here due to the single-path assumption.
\begin{table*}[t] \centering
\caption{Simulation assumptions and parameters.}\label{tracking_comp}
\begin{tabular}{||c||c||}
  \hline
  \textbf{SYSTEM PARAMETERS} & \textbf{SIMULATION ASSUMPTIONS} \\
  \hline
  BS-UE distance $d$ (m) & $100$ \\
  \hline
  UE's azimuth velocity $v$/$v_{\mathrm{az}}$ (km/h) & $100$ \\
  \hline
  UE's elevation velocity $v_{\mathrm{el}}$ (km/h) & $30$ \\
  \hline
  Symbol duration ($\mu s$) & $3.7$ \\
  \hline
  Total number of symbols $T_{\mathrm{tot}}$ & $10^4$ \\
  \hline
  Periodicity of the DTC $T_{\mathrm{d}}$ (number of symbols) & $10$, $100$, $1000$, $2000$ \\
  \hline
  Azimuth angle variation model & $\psi_{t} = \psi_{t-1}+\Delta\psi + w$ \\
  \hline
  Elevation angle variation model & $\theta_{t} = \theta_{t-1}+\Delta\theta + w$ \\
  \hline
  Pathloss model & \cite{kssr} \\
  \hline
\end{tabular}
\end{table*}
Note that the angle variations $\Delta\psi$ and $\Delta\theta$ are obtained according to the UE's azimuth and elevation velocities $v_{\mathrm{az}}$ and $v_{\mathrm{el}}$, the BS-UE distance $d$, and the symbol duration. We further randomize the angle variations by incorporating a Gaussian distributed random variable $w$ with zero mean and variance $1$ as in Table I. In the simulations, we set $T=1$. That is, each DTC comprises one time-slot (symbol), during which one auxiliary beam pair is formed. The two beams in the corresponding auxiliary beam pair are simultaneously probed, which are differentiated by the UE via the beam-specific pilot signal design. We can then define the tracking overhead as $\rho=1/T_{\mathrm{d}}$. For instance, $T_{\mathrm{d}}=1000$ results in less tracking overhead than $T_{\mathrm{d}}=10$ as the corresponding tracking overheads are computed as $\rho=0.1\%$ and $\rho=10\%$. We assume angular motion model I for Figs.~8 and 9, and angular motion model II for Fig.~10.

\begin{figure}
\centering
\subfigure[]{%
\includegraphics[width=2.78in]{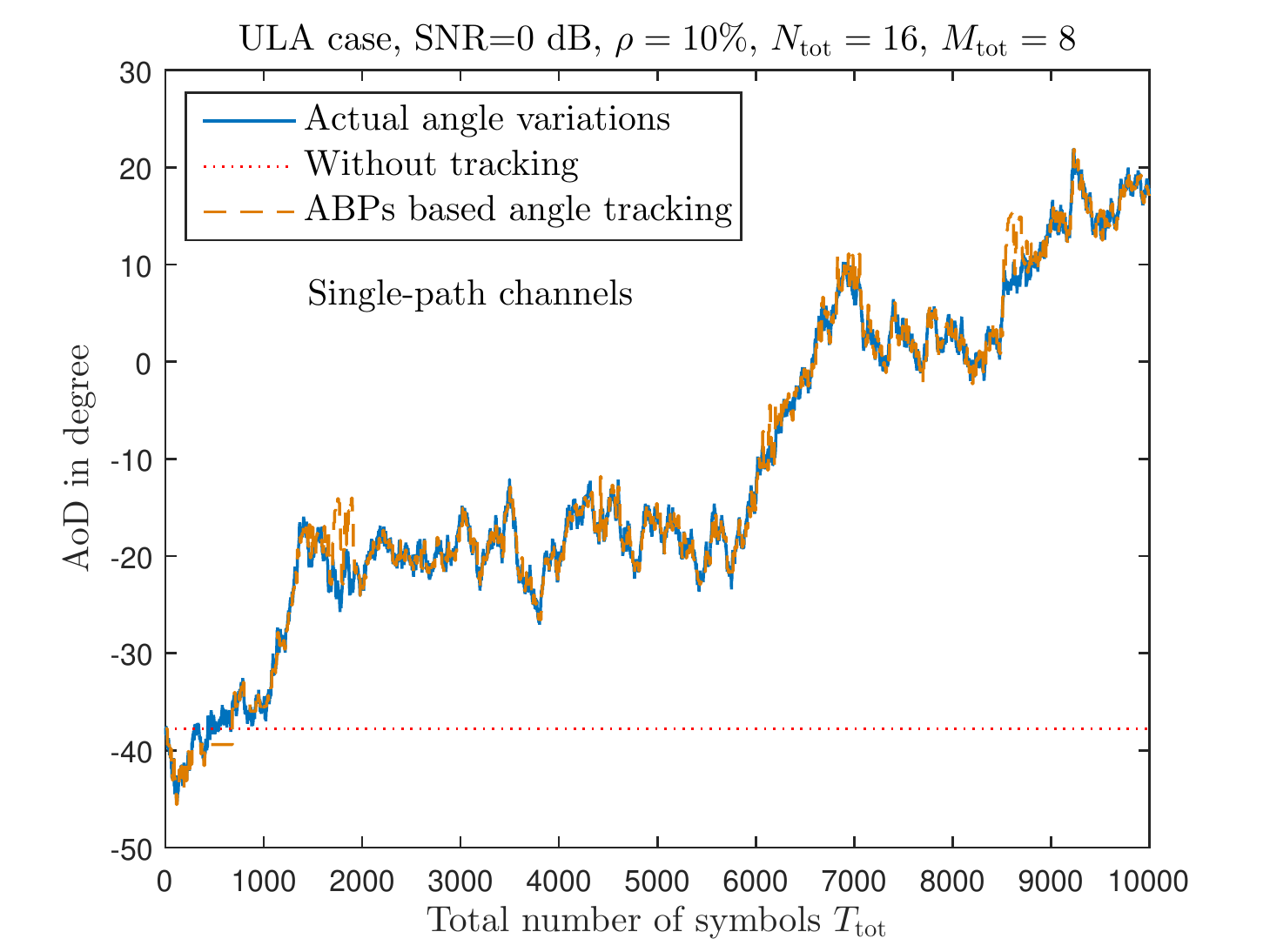}
\label{fig:subfigure1}}
\subfigure[]{%
\includegraphics[width=2.78in]{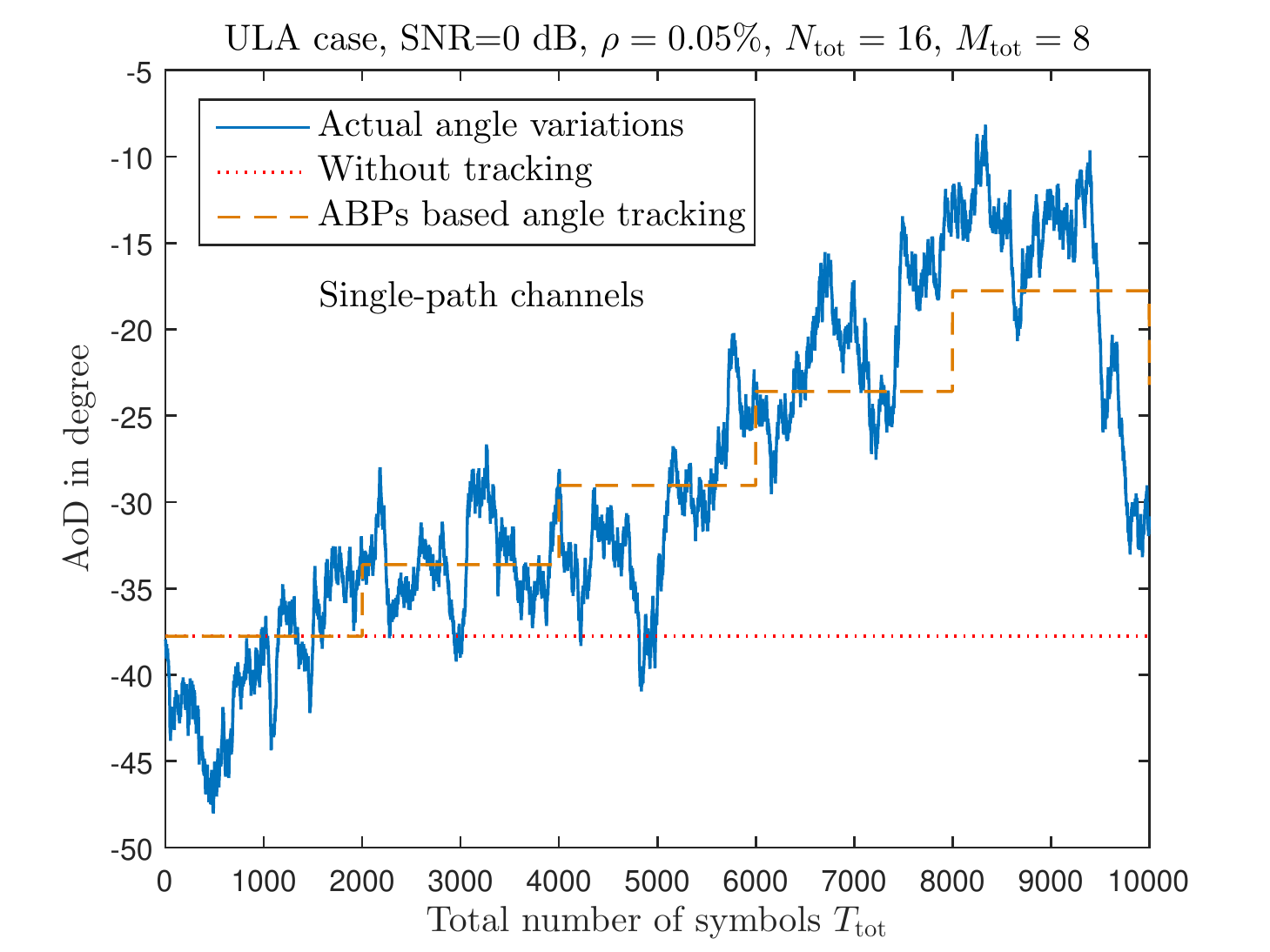}
\label{fig:subfigure2}}
\caption{Examples of actual angle variations, angle tracking using the proposed method, and without angle tracking in single-path channels; $N_{\mathrm{tot}}=16$, $M_{\mathrm{tot}}=8$ and $0$ dB SNR are assumed with the ULA equipped at the BS and ideal radiation pattern. (a) $\rho=10\%$ tracking overhead. (b) $\rho=0.05\%$ tracking overhead.}
\label{fig:figure}
\end{figure}
In Fig.~8, we provide snapshots of the angle tracking results over time for $\rho=10\%$ and $0.05\%$ in the proposed design. For comparison, we also provide the actual angle variations and the case without angle tracking. Further, we assume ideal radiation pattern. As can be seen from Fig.~8(a), the proposed auxiliary beam pair-assisted angle tracking design can accurately track the angle variations under relatively high tracking overheads, i.e., $10\%$. By reducing the tracking overhead to $0.05\%$, the tracking resolution becomes small, which in turn, degrades the angle tracking performance as shown in Fig.~8(b). Under different tracking overheads assumptions, the trend of the angle variations can be well captured by employing the proposed angle tracking design.

\begin{figure}
\centering
\subfigure[]{%
\includegraphics[width=2.78in]{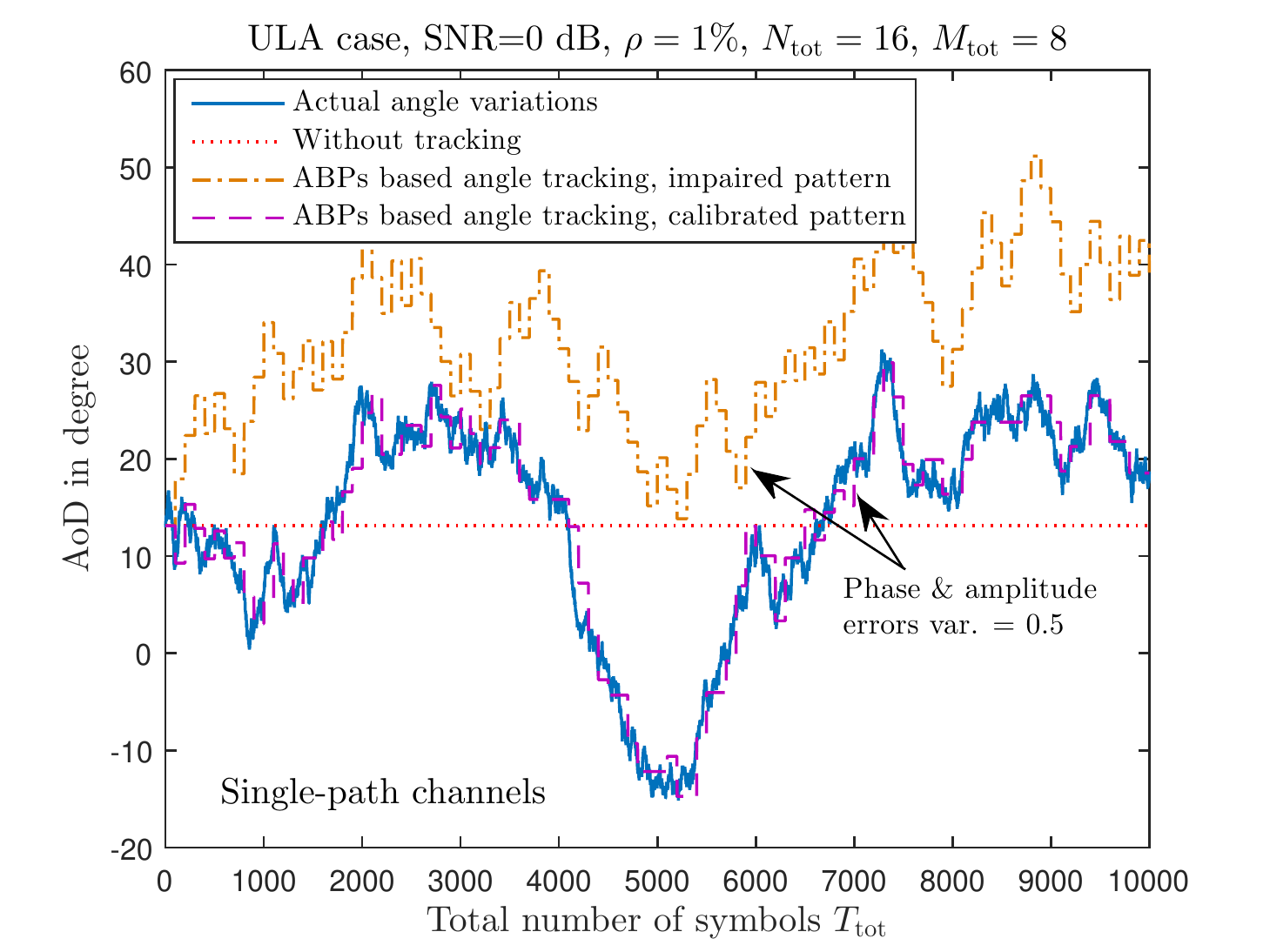}
\label{fig:subfigure1}}
\subfigure[]{%
\includegraphics[width=2.78in]{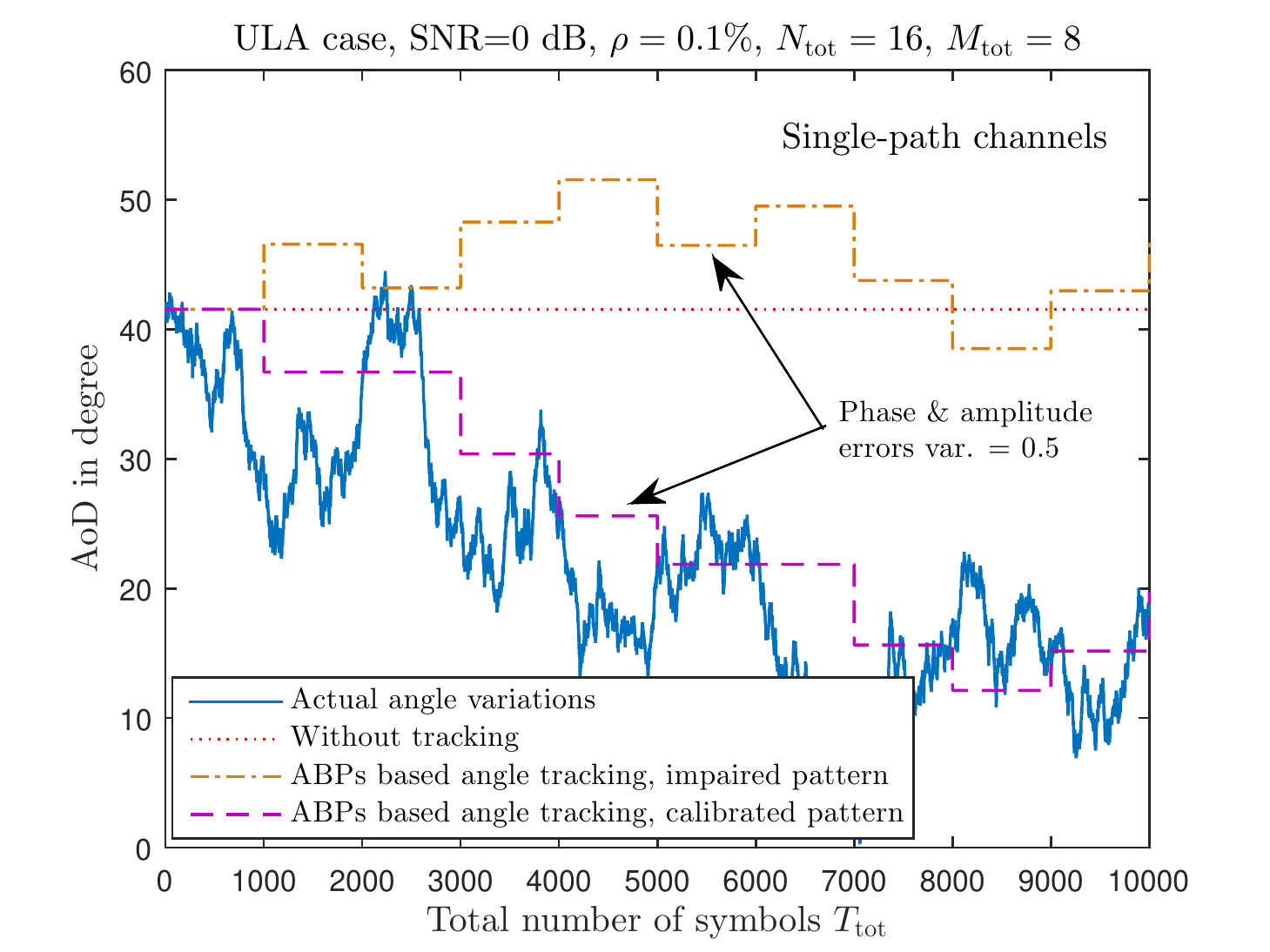}
\label{fig:subfigure2}}
\caption{$N_{\mathrm{tot}}=16$, $M_{\mathrm{tot}}=8$ and $0$ dB SNR are assumed with the ULA equipped at the BS; ideal radiation pattern, impaired radiation pattern, and calibrated radiation pattern are evaluated. The phase $\&$ amplitude errors variances are set as $0.5$. (a) Examples of actual angle variations, angle tracking using the proposed method, and without angle tracking; $\rho=1\%$ tracking overhead. (b) Examples of actual angle variations, angle tracking using the proposed method, and without angle tracking; $\rho=0.1\%$ tracking overhead.}
\label{fig:figure}
\end{figure}
In Figs.~9(a) and (b), we evaluate the proposed angle tracking design assuming both impaired and calibrated radiation patterns with $1\%$ and $0.1\%$ tracking overheads. We set the variances of the phase and amplitude errors as $0.5$. Due to the random phase and amplitude errors, the angle tracking performance of the proposed approach with the impaired radiation pattern is deteriorated. Even with relatively high tracking overheads (e.g., $1\%$ in Fig.~9(a)), the tracked angles are very different from the actual ones for all the channel realizations. By compensating for the phase and amplitude errors via the proposed calibration method, the angle tracking performance is significantly improved. For $1\%$ tracking overhead, the angle tracking performance of the proposed method with calibrated radiation pattern almost matches with the actual angle variations for all the channel realizations.



\begin{figure}
\centering
\subfigure[]{%
\includegraphics[width=2.78in]{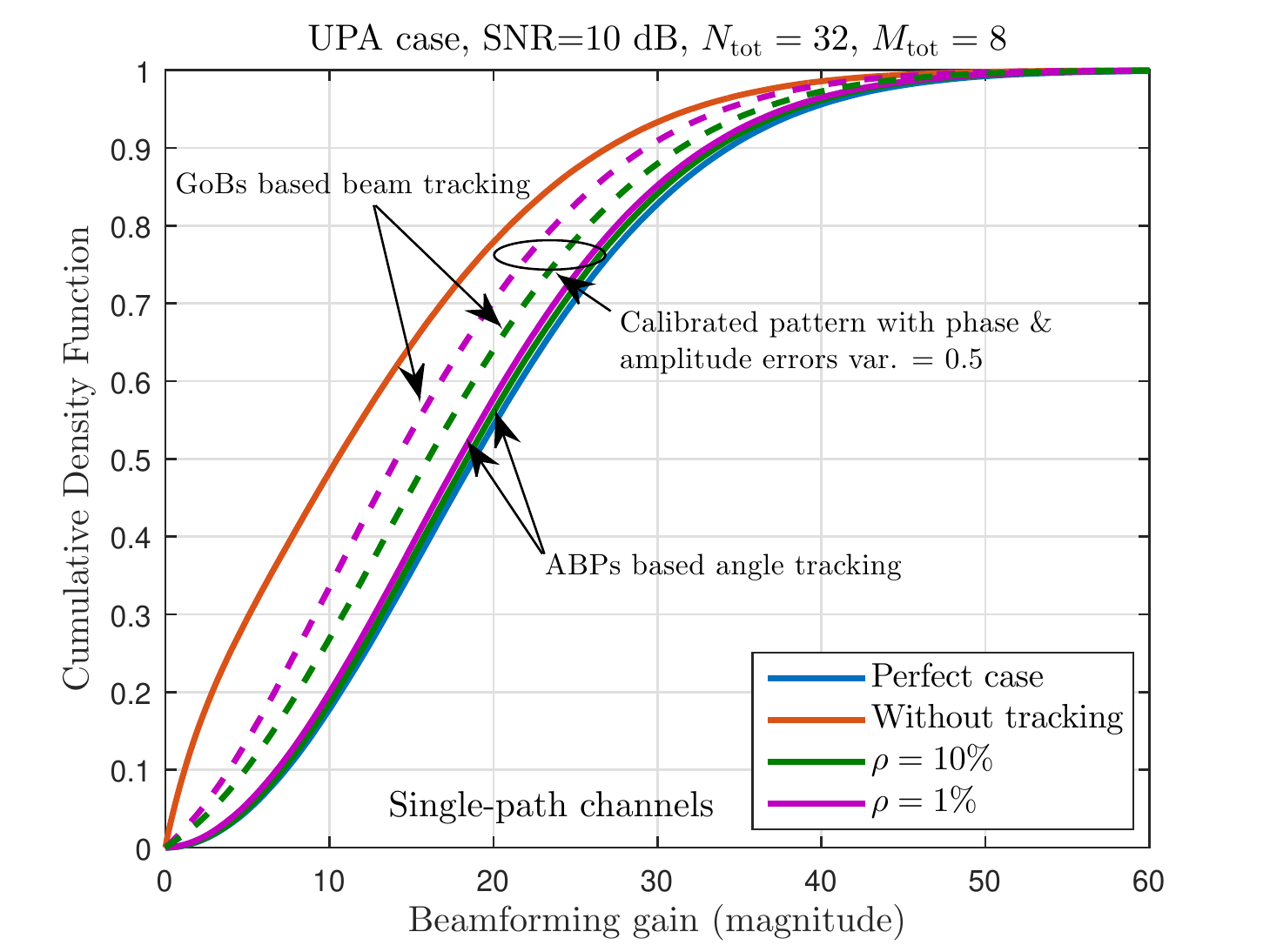}
\label{fig:subfigure1}}
\subfigure[]{%
\includegraphics[width=2.78in]{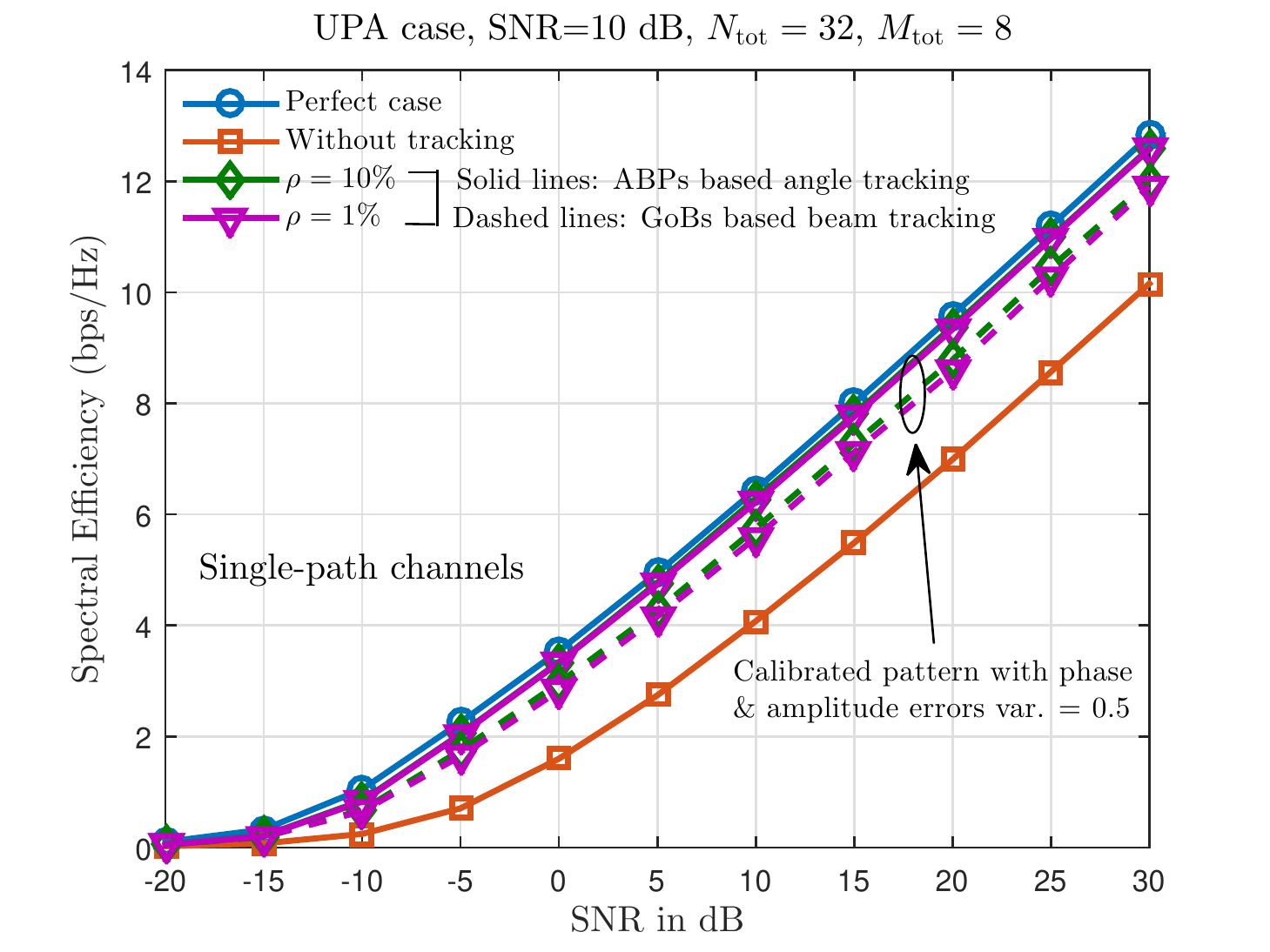}
\label{fig:subfigure2}}
\caption{Angular motion model II. (a) CDFs of the beamforming gains obtained via the anchor beams in the DDC with $N_{\mathrm{tot}}=32$, $M_{\mathrm{tot}}=8$ and $10$ dB SNR. The proposed approach and the grid-of-beams (GoBs) based beam tracking are evaluated assuming calibrated radiation pattern. (b) Spectral efficiency performance obtained via the anchor beams in the DDC with $N_{\mathrm{tot}}=32$, $M_{\mathrm{tot}}=8$ and $10$ dB SNR. The proposed approach and the grid-of-beams based beam tracking are evaluated assuming calibrated radiation pattern.}
\label{fig:figure}
\end{figure}
We now evaluate the two-dimensional angle tracking performance for the proposed approach using calibrated radiation pattern. A total of $N_{\mathrm{tot}}=32$ antenna elements are equipped at the BS side with the UPA placed in the $\mathrm{xy}$-plane. Further, we set $N_{\mathrm{x}}=4$ and $N_{\mathrm{y}}=8$. In Fig.~10(a), we plot the cumulative density functions (CDFs) of the beamforming gains obtained from the anchor beam in the DDC. With calibrated radiation pattern, the proposed method shows close performance relative to the perfect case assuming various tracking overheads. In Fig.~10(b), the spectral efficiency performance is evaluated using the anchor beam in the DDC. Specifically, denoting by $h_{\mathrm{eff}} = g\bm{a}_{\mathrm{r}}^{*}(\nu)\bm{a}_{\mathrm{r}}(\nu)\bm{a}_{\mathrm{t}}^{*}(\theta,\psi)\bm{a}_{\mathrm{t}}(\hat{\theta},\hat{\psi})$ for single-path channels, we can compute the spectral efficiency metric as $C = \mathbb{E}\left[\log_{2}\left(1+\gamma h_{\mathrm{eff}}^{*}h_{\mathrm{eff}}\right)\right]$. Similar to Fig.~10(a), the spectral efficiency performances obtained by using the proposed method with different tracking overheads are close to the perfect case. In Figs.~10(a) and (b), we also evaluate the grid-of-beams based beam tracking design assuming various tracking overheads. For fair comparison, we employ the same number of tracking beams as in the auxiliary beam pair based angle tracking design. As can be seen from Figs.~10(a) and (b), the proposed algorithm shows superior beamforming gain and spectral efficiency performances over the grid-of-beams based beam tracking strategy.
\subsection{Wideband multi-path channels with OFDM}
\begin{figure}
\centering
\subfigure[]{%
\includegraphics[width=4.8in]{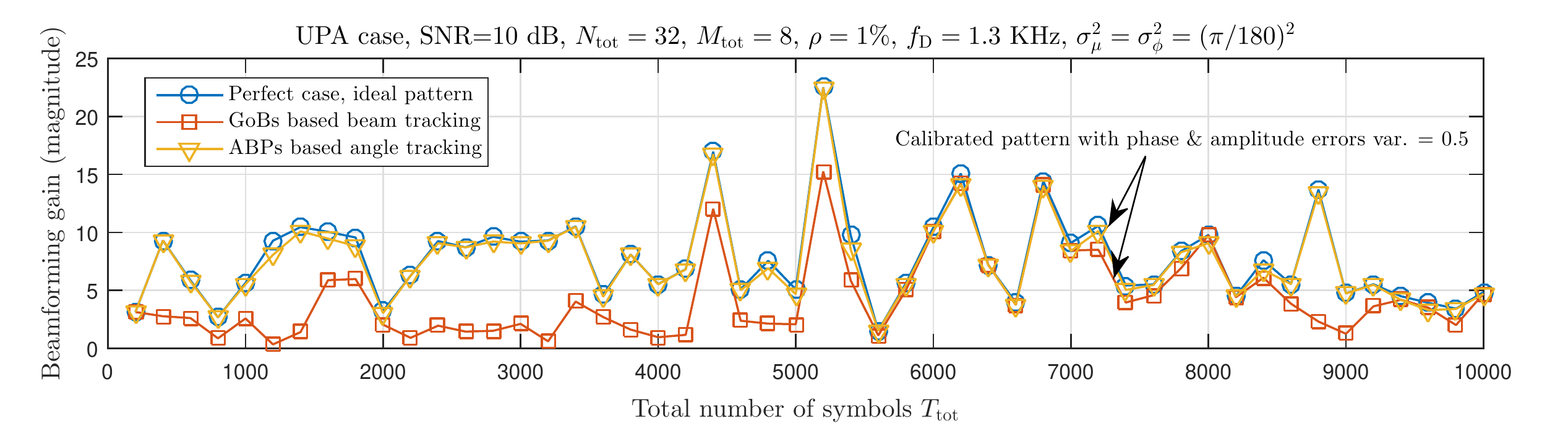}
\label{fig:subfigure1}}
\subfigure[]{%
\includegraphics[width=4.8in]{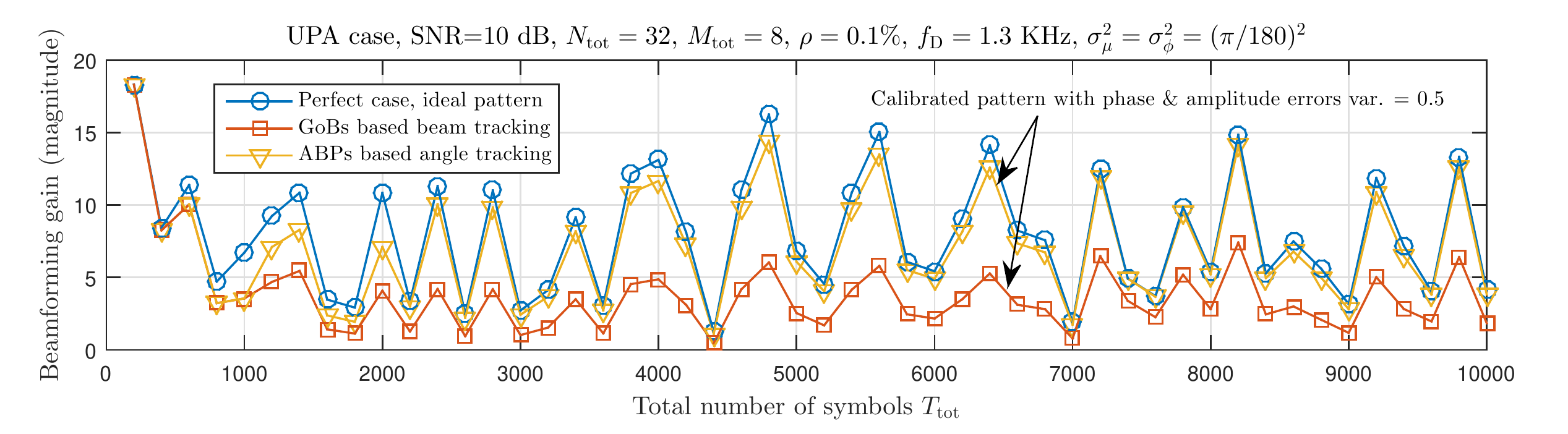}
\label{fig:subfigure2}}
\caption{Beamforming gains obtained via the anchor beams in the DDC with $N_{\mathrm{tot}}=32$, $M_{\mathrm{tot}}=8$ and $10$ dB SNR. The proposed approach and the grid-of-beams based beam tracking are evaluated assuming calibrated radiation pattern. $f_{\mathrm{D}}=1.3$ KHz and $\sigma^{2}_{\mu}=\sigma^{2}_{\phi}=(\pi/180)^2$. (a) $\rho=1\%$ tracking overhead. (b) $\rho=0.1\%$ tracking overhead.}
\label{fig:figure}
\end{figure}
The temporal evolution effect of mmWave channels is not well characterized in current wideband mmWave channel models \cite{timemmwave}. In this part of simulation, we therefore first implement the temporally correlated mmWave channels by considering both (i) the NYUSIM open source platform developed in \cite{nyuweb} and (ii) the statistical temporal evolution model used in \cite{bt5,trackingsystem}.

For the NYUSIM open source platform, we consider the urban micro-cellular (UMi) scenario with NLOS components for the $28$ GHz carrier frequency. We evaluate $125$ MHz RF bandwidths with $N=512$ subcarriers. The corresponding CP lengths is $D=64$. The employed ZC-type sequences occupy the central $63$ subcarriers with the root indices $i_0=25$ and $i_1=34$. We set the subcarrier spacing and symbol duration as $270$ KHz and $3.7$ $\mu s$ following the numerology provided in \cite{jerrypikhan}. Detailed channel modeling parameters are given in \cite[Table~III]{kssr}. Further, our design focus here is to track the strongest path's AoD by using the proposed approach.

Before proceeding with the temporal channel evolution model, we first rewrite the time-domain channel matrix in (\ref{delayd}) in a more compact form. For time-slot $t$, denoting by $\bm{\varphi}_{t}=\left[\varphi_{1,t},\varphi_{2,t},\cdots,\varphi_{N_{\mathrm{r}},t}\right]^{\mathrm{T}}$, $\bm{\mu}_{t}=\left[\mu_{1,t},\mu_{2,t},\cdots,\mu_{N_{\mathrm{r}},t}\right]^{\mathrm{T}}$ and $\bm{\phi}_{t}=\big[\phi_{1,t},\phi_{2,t},\cdots,\phi_{N_{\mathrm{r}},t}\big]^{\mathrm{T}}$, we have
\begin{equation}
\bm{H}_{t}[d] = \bm{A}_{\mathrm{R}}(\bm{\varphi}_{t})\bm{G}_{t}[d]\bm{A}^{*}_{\mathrm{T}}(\bm{\mu}_{t},\bm{\phi}_{t}),
\end{equation}
where $\bm{A}_{\mathrm{R}}(\bm{\varphi}_{t})$ and $\bm{A}_{\mathrm{T}}(\bm{\mu}_{t},\bm{\phi}_{t})$ represent the array response matrices for the receiver and transmitter such that
\begin{eqnarray}
&\bm{A}_{\mathrm{R}}(\bm{\varphi}_{t}) = \left[\bm{a}_{\mathrm{r}}(\varphi_{1,t})\hspace{3mm}\bm{a}_{\mathrm{r}}(\varphi_{2,t})\hspace{3mm}\cdots\hspace{3mm}\bm{a}_{\mathrm{r}}(\varphi_{N_{\mathrm{r}},t})\right]&\\
&\bm{A}_{\mathrm{T}}(\bm{\mu}_{t},\bm{\phi}_{t}) = \left[\bm{a}_{\mathrm{t}}(\mu_{1,t},\phi_{1,t})\hspace{3mm}\bm{a}_{\mathrm{t}}(\mu_{2,t},\phi_{2,t})\hspace{3mm}\cdots\hspace{3mm}\bm{a}_{\mathrm{t}}(\mu_{N_{\mathrm{r}},t},\phi_{N_{\mathrm{r}},t})\right],&
\end{eqnarray}
and $\bm{G}_{t}[d]=\mathrm{diag}\left(\left[g_{1}p\left(dT_{\mathrm{s}}-\tau_1\right),\cdots,g_{N_{\mathrm{r}}}p\left(dT_{\mathrm{s}}-\tau_{N_{\mathrm{r}}}\right)\right]^{\mathrm{T}}\right)$. We model the temporal evolution of the path gains as the first-order Gauss-Markov process as \cite{bt5}
\begin{equation}\label{gddd}
\bm{G}_{t+1}[d] = \rho_{\mathrm{D}}\bm{G}_{t}[d]+\sqrt{1-\rho_{\mathrm{D}}^2}\bm{B}_{t+1},
\end{equation}
where $\rho_{\mathrm{D}}=J_0\left(2\pi f_{\mathrm{D}}T_{\mathrm{s}}\right)$ and $\bm{B}_{t+1}$ is a diagonal matrix with the diagonal entries distributed according to $\mathcal{N}_{c}(0,1)$. Here, $J_0(\cdot)$ denotes the zeroth-order Bessel function of first kind and $f_{\mathrm{D}}$ is the maximum Doppler frequency. The elevation and azimuth AoDs vary according to \cite{trackingsystem}
\begin{equation}\label{addd}
\bm{\mu}_{t+1} = \bm{\mu}_{t}+\Delta\bm{\mu}_{t+1},\hspace{3mm}\bm{\phi}_{t+1} = \bm{\phi}_{t}+\Delta\bm{\phi}_{t+1},
\end{equation}
where $\Delta\bm{\mu}_{t+1}$ and $\Delta\bm{\phi}_{t+1}$ are distributed according to $\mathcal{N}_{c}(\bm{0}_{N_{\mathrm{r}}},\sigma^{2}_{\mu}\bm{I}_{N_{\mathrm{r}}})$ and $\mathcal{N}_{c}(\bm{0}_{N_{\mathrm{r}}},\sigma^{2}_{\phi}\bm{I}_{N_{\mathrm{r}}})$. We first determine the initial path gains, path delays, azimuth/elevation AoDs, and AoAs through one simulation run using the NYUSIM open source platform. We then obtain the channels for the subsequent time-slots by using the initial channel results and the temporal evolution model presented in (\ref{gddd}) and (\ref{addd}).

In Fig.~11, we plot the beamforming gains against the employed OFDM symbols for $\rho=1\%$ and $0.1\%$ tracking overheads. We set $f_{\mathrm{D}}=1.3$ KHz and $\sigma^{2}_{\mu}=\sigma^{2}_{\phi}=(\pi/180)^{2}$, which characterize relatively fast moving and angle variation speeds \cite{bt5,trackingsystem}. In addition to the actual angle variations, we evaluate the proposed angle tracking and grid-of-beams based beam tracking designs with calibrated radiation patterns. Similar to the evaluation results shown in Section V-A, the proposed algorithm shows close tracking performance to the perfect case, and outperforms the existing beam tracking approach for various system setups.
\section{Conclusions}
In this paper, we developed and evaluated several new angle tracking design approaches for mobile wideband mmWave systems with antenna array calibration. The proposed methods are different in terms of tracking triggering metric, feedback information, and auxiliary beam pair setup required at the UE. These differences allow the proposed strategies to be adopted in different deployment scenarios. We exposed the detailed design procedures of the proposed methods and showed that they can obtain high-resolution angle tracking results. The proposed methods neither depend on a particular angle variation model nor require the on-grid assumption. Since the proposed methods are sensitive to radiation pattern impairments, we showed by numerical examples that with appropriate array calibration, the angle variations can still be successfully tracked via the proposed methods under various angle variation models.

\bibliographystyle{IEEEbib}
\bibliography{main_TWC}

\end{document}